\RequirePackage[T1]{fontenc}
\documentclass[12pt]{article}

\usepackage{comment}
\usepackage{titling}

\usepackage{authblk}
\usepackage{mathtools}

\usepackage{graphicx}
\makeatletter
\let\oldbullet\bullet
\renewcommand{\bullet}{%
  \mathbin{\vcenter{\hbox{\scalebox{1.2}{$\m@th\oldbullet$}}}}% 
}
\makeatother

\usepackage{times}

\usepackage[utf8]{inputenc}
\usepackage{amsmath}
\usepackage{amssymb}
\usepackage{mathtools}
\numberwithin{equation}{section}
\usepackage{slashed}
\usepackage{braket}
\usepackage{bm}
\usepackage{mathrsfs}
\usepackage[svgnames]{xcolor}

\usepackage[colorlinks,citecolor=DarkGreen,linkcolor=FireBrick,linktocpage,pagebackref]{hyperref}
\usepackage{cite}

\usepackage{caption}

\usepackage{times}

\usepackage[utf8]{inputenc}

\usepackage{here}
\usepackage[normalem]{ulem}%取り消し線

\usepackage{url}

\usepackage{dsfont}

\usepackage[top=30truemm,bottom=30truemm,left=25truemm,right=25truemm]{geometry}

\usepackage{spectralsequences}
\usepackage{tikz}
\usetikzlibrary{intersections}
\usetikzlibrary{calc}

\usepackage{xcolor}
\usepackage{tikz-cd}

\usepackage{slashed}

\usepackage{amsthm}
\usepackage{thmtools}
\usepackage{blindtext}
\usepackage{braket}

\declaretheoremstyle[
       shaded={bgcolor=\color{rgb}{0.9,0.9,0.9}}  % comment this line in/out
]{theorem}

\declaretheoremstyle[
       shaded={bgcolor=\color{rgb}{0.9,0.9,0.9}}% comment this line in/out
]{question}

\declaretheoremstyle[
       shaded={bgcolor=\color{rgb}{0.9,0.9,0.9}}  % comment this line in/out
]{remark}

\declaretheoremstyle[
       shaded={bgcolor=\color{rgb}{0.9,0.9,0.9}}  % comment this line in/out
]{proposition}

\declaretheoremstyle[
       shaded={bgcolor=\color{rgb}{0.9,0.9,0.9}}  % comment this line in/out
]{definition}

\declaretheoremstyle[
       shaded={bgcolor=\color{rgb}{0.9,0.9,0.9}}  % comment this line in/out
]{assumption}

\declaretheoremstyle[
       shaded={bgcolor=\color{rgb}{0.9,0.9,0.9}}  % comment this line in/out
]{conjecture}

\declaretheoremstyle[
       shaded={bgcolor=\color{rgb}{0.9,0.9,0.9}}  % comment this line in/out
]{corrorary}

\declaretheoremstyle[
       shaded={bgcolor=\color{rgb}{0.9,0.9,0.9}}  % comment this line in/out
]{axiom}

\declaretheoremstyle[
       shaded={bgcolor=\color{rgb}{0.9,0.9,0.9}}  % comment this line in/out
]{lemma}

\usepackage{mathrsfs}
\usepackage{amsmath,amssymb}

\usepackage{esvect}

\usepackage{adjustbox}

\usepackage[status=draft,inline,nomargin]{fixme}
\fxusetheme{color}
\FXRegisterAuthor{ki}{aki}{KI}
\FXRegisterAuthor{shu}{shuenv}{\color{red}Shu}

\usetikzlibrary{decorations.pathmorphing,shapes}
\newcounter{sarrow}

\usepackage{xurl}

\def\[#1\]{%
  \begin{equation}\begin{gathered}#1\end{gathered}\end{equation}%
}

\begin{document}

\title{Parameterized Families of Toric Code Phase:\\ $em$-duality family and higher-order anyon pumping}

\author[1,2]{Shuhei Ohyama}
\author[3]{Takamasa Ando}
\author[4]{Ryan Thorngren}

\affil[1]{University of Vienna, Faculty of Physics, Boltzmanngasse 5, A-1090, Vienna, Austria}
\affil[2]{
RIKEN Center for Emergent Matter Science, Wako, Saitama, 351-0198, Japan}
\affil[3]{Center for Gravitational Physics and Quantum Information, Yukawa Institute for Theoretical Physics, Kyoto University, Kyoto 606-8502, Japan}
\affil[4]{Mani L. Bhaumik Institute for Theoretical Physics, Department of Physics and Astronomy, University of California, Los Angeles, CA 90095, USA}

\date{} 
% \preprint{YITP-25-124}

\maketitle

\begin{abstract}
Within the toric-code phase, we study parameterized families of topologically ordered states. We construct $1$- and $2$-parameter families of local Hamiltonians and confirm their non-triviality via topological pumping. For the $1$-parameter family, we show that the $em$-exchange defect is pumped into the bond Hilbert space of a tensor-network representation. For the $2$-parameter case, we construct a ``pump of a pump'' that transports an 
$S^1$-family of a system in one lower spatial dimension. Using similar methods, we also present a $1$-parameter family with a higher-order anyon pump that produces corner-localized anyon modes. These constructions provide explicit lattice realizations and concrete diagnostics of family-level topology. We use recently developed boundary algebra methods to study the non-triviality of these families.
\end{abstract}

\setcounter{tocdepth}{3}
\tableofcontents

\section{Introduction and Organization}
% \label{sec: Introduction and Summary}

\paragraph{Introduction}

Recently, there has been a lot of progress studying adiabatic quantized invariants of systems with a topologically non-trivial parameter space $X$. For example, if we consider a single periodic control parameter, $X$ is a circle $S^1$, and we may study what happens as this parameter varies adiabatically around the cycle. Famously, in an $S^1$-family of trivial\footnote{Trivial means states which are in the same phase as a product state.} gapped states with a conserved $U(1)$ charge, if we vary the parameter $\theta$, there may be a current $J^x \propto \partial_t \theta$ to leading order in the expansion in $\partial_t \theta$. This results in a ``Thouless charge pump'' around the cycle. In fact, completing a full cycle always must result in an integer amount $q$ of charge which is pumped, so $J^x = \frac{q}{2\pi} \partial_t \theta$ with quantized coefficient \cite{Thouless83,niu1984quantised}.

We can regard this as a topological invariant of the $S^1$-family. More precisely, if the pumped charge is non-zero, the $S^1$-family is ``non-contractible'', meaning that it does not extend to a symmetric gapped family over a disc $D^2$. In other words, suppose we enlarge our parameter space to $D^2$, such that our original parameter $\theta$ is the angular coordinate on $D^2$, and the radial coordinate $r$ is a new control parameter. This is very natural if we are thinking about a periodically driven system, where $r$ may be a measure of the strength of the drive. When the strength of the drive goes to zero there is no longer any sensitivity to its phase. Non-contractibility of the $S^1$-family means the following. Suppose there is a non-zero charge pumped as we vary $\theta$ at strength $r = 1$. As we decrease the strength, eventually the charge which is pumped must be zero at $r = 0$, since the system is now static. This means at some $r$ the adiabatic approximation must break down, implying that there is some kind of singularity in the phase diagram of the system as a function of $(r,\theta)$, such as an intervening phase at small $r$ or an isolated gapless point known as a diabolical point, whose study is of independent interest \cite{Teo:2010zb,Cordova:2019jnf,Cordova:2019uob,Hsin:2020cgg,Tantivasadakarn:2021wdv,Jones:2024zhx,Prakash:2024yfr,Jones:2025khc,Manjunath:2026ezp}.

The Thouless charge pump has been generalized to $S^1$-families of higher dimensional familes of $G$-symmetric trivial states, which pump a $G$-symmetry protected topological (SPT) phase to their boundary over an adiabatic cycle \cite{kitaev2015homotopy,Else:2016hyb,roy2017floquet,Tantivasadakarn:2021noi,Inamura:2026hjl}. One may also generalize the parameter space $X$, for example to higher dimensional spheres, where one finds pumping invariants associated to parameters varying slowly in space and time \cite{Kapustin:2020mkl,Shiozaki:2021weu,Artymowicz:2023erv,Ohyama:2026oay}. These can be understood as charges (or entire $G$-SPTs) bound to skyrmion-like configurations wrapping cycles of $X$ \cite{Hsin:2020cgg}. In the absence of symmetry there is a also many-body ``higher Berry phase''\cite{Kitaev_Freed60,Kapustin:2020eby,Wen:2021gwc,Ohyama:2023suc,OR23,Qi:2023ysw,Spiegel:2023lhv,shiozaki2023higher,Ohyama:2024jsg,Ohyama:2024ytt,Sommer:2024dtb,Sommer:2024lzp,Liu:2024ulq,Artymowicz:2023erv,Geiko:2024cra,Spiegel:2025ugu,wen2025spaceconformalboundaryconditions,Shiozaki:2025pyo}, which occurs as a quantized ``Wess-Zumino-Witten term'' for the effective theory of the slowly varying parameters\cite{Hsin:2020cgg,Choi:2022odr,Choi:2025ebk}.

The above examples all occur in the trivial gapped phase.\footnote{
    The higher Berry curvature for general gapped phases is defined in \cite{Kapustin:2020eby,Kapustin:2022apy}.
} If we consider non-trivial gapped phases, especially topologically ordered states, there are new phenomena. The goal of this paper is to construct and study some simple models of some of these new phenomena, particularly in two spatial dimensions, which is an instructive setting.

Topologically ordered systems in two spatial dimensions are gapped quantum systems characterized by the presence of anyons, which are quasiparticles that exhibit nontrivial braiding statistics.
Such systems were mainly discovered in the research on the fractional quantum Hall effect\cite{PhysRevLett.48.1559,Wen:1989iv,MooreRead1991}, but today they also play an important role in quantum computation, for example as quantum error correcting codes\cite{Freedman:2000gwh,Kitaev_2003,freedman2002topologicalquantumcomputation}. The algebraic structure of anyons is mathematically described by modular tensor categories \cite{BakalovKirillov,Turaev2016}.

It is believed that families of topologically ordered states parametrized by spheres $X = S^k$ have a simple structure \cite{Kitaev_2006,Aasen:2022cdu,Hsin:2022iug}. If we consider a ``Skyrmion'' configuration where the parameters wrap the sphere, this defines a codimension-$k$ defect in the topological order. The superselection sector of this defect is a quantized invariant of the family. The result is that in two spatial dimensions we will have
\begin{enumerate}
    \item $S^1$-families labeled by invertible anyon-permuting and more general symmetries of the topological order, corresponding to invertible codimension 1 defects
    \item $S^2$-families labeled by (abelian) anyons
    \item $S^3$-families have no quantized labels native to the topological order because there are no point-like (in spacetime) superselection sectors
\end{enumerate}
as well as versions of the families in trivial gapped states, such as a quantized higher Berry-Chern number for $S^4$-families. Note these match the homotopy groups of the $3$-group defined in \cite{ENO2010}. This $3$-group is thought to be a kind of approximation to the ``moduli space'' of 2+1D topologically ordered states \cite{Hsin:2022iug}.\footnote{More precisely, it is the moduli space of non-chiral topological orders. In this paper, we consider only non-chiral topologically ordered states.} In general $X$-families of topologically ordered states should be classified by homotopy classes of maps from $X$ into this moduli space.

In the case of the toric code, which has bosonic $\mathbb{Z}_2$ anyons $e$ and $m$ with a mutual $-1$ braiding, we should thus have
\begin{enumerate}
    \item A non-trivial $S^1$-family associated with the $e$-$m$ swap symmetry.
    \item Three non-trivial $S^2$-families associated with each of $e$, $m$, and their fusion $em$.
\end{enumerate}
We construct toric code families of the first type in Section~\ref{sec: em pump}, as well as two-parameter families similar to the second type in Sections \ref{sec: 2-para pump II} and \ref{sec: crystalline pump}.\footnote{
More precisely, what we consider in Section~\ref{sec: crystalline pump} is an $S^{1}$-family with crystalline symmetry.
}

To achieve this, we use a ``symmetry interpolation'' method which is quite general. Suppose a Hamiltonian $H$ possesses a unitary symmetry $U$, and that we can find an interpolation $U(\theta)$ from the identity $U(0) = I$ to $U(2\pi) = U$. Then
\begin{equation}\label{eqnsymmetryinterp}
    H(\theta) \coloneqq U(\theta) H U(\theta)^\dagger
\end{equation}
is a family of Hamiltonians which is $2\pi$-periodic in $\theta$. Furthermore, all these Hamiltonians have the same spectra, so if $H$ is gapped, they all are. We see on the other hand that while $H$ is periodic, under adiabatic evolution, any eigenstate $\ket{\psi}$ evolves to $U\ket{\psi}$, thus implementing the symmetry.

One challenge therefore is to find a realization $U_{em}$ of the $e$-$m$ swap symmetry which one may interpolate. The most famous realization of this symmetry involves a translation of the lattice and cannot be interpolated \cite{Dennis_2002}. It would be best to have an on-site realization, where $U(\theta)$ is a product of single-site unitaries which could then be interpolated. On-site realizations of $e$-$m$ swap exist in string-net models in the same phase\footnote{This gives an easy construction of an $e$-$m$ swapping $S^1$-family in the toric code phase. See also \cite{Aasen:2022cdu,Po_2017} for other constructions in the toric code phase.} \cite{Heinrich:2016wld}, but no such realization exists (or indeed can exist) for the toric code model itself \cite{ma2024quantumcellularautomatasymmetric,Jones_2026}. However, constant depth circuit realizations of this symmetry in the toric code exist (see Section~3.2 of \cite{Barkeshli:2022wuz}) and have recently been revisited in \cite{Tu:2025bqf,Shirley:2025yji} for their connection to the study of anomalies on the lattice. We use such a circuit to define an interpolated $S^1$-family starting with the toric code.

By construction, this $S^1$-family has the property that $e$ and $m$ anyons exchange their identity after advancing the parameter $\theta$ by $2\pi$. This strongly suggests that the family is non-contractible in the sense defined above. Restricting to unitary families of the type \eqref{eqnsymmetryinterp}, we develop a bulk-boundary correspondence that allows us to show non-contractibility in this setting. That is, we will argue that there is no family of finite-depth unitaries $U(r,\theta)$ with $U(1,\theta) = U(\theta)$ and $U(0,\theta) = I$.

The bulk-boundary correspondence is based on studying the boundary algebra \cite{jones2023localtopologicalorderboundary}. Intuitively, the boundary algebra is the algebra of local operators on the boundary of a topologically ordered state. It can be defined for any ground state subspace which is given by a ``net of projections'' satisfying some local topological order axioms which are satisfied for example by the toric code ground state projections (see Section~2 of \cite{jones2023localtopologicalorderboundary} for precise definitions). They will also be satisfied for the Hamiltonians $H(\theta)$ in the unitary family \eqref{eqnsymmetryinterp}, and so we can define a bundle of boundary algebras fibered over the parameter space. For an $S^1$-family, this bundle is characterized by an automorphism of the algebra, which comes from the symmetry $U$ which is interpolated. We compute the automorphism and show it carries a non-trivial index of the type defined in \cite{Jones:2023imy}. This index does not admit continuous deformations, so the family is non-contractible.

We will show this automorphism permutes the DHR bimodules of the boundary algebra\cite{Jones:2023ptg,jones2023localtopologicalorderboundary,Jones:2023imy,Jones_2026,Hataishi2025,Jones:2025yhx,Jones:2026dcb}, corresponding to the swapping of $e$ and $m$ superselection sectors, which are identified with these bimodules. We provide a simple and intuitive description of these bimodules which makes this clear. For more general gapped families not of the form \eqref{eqnsymmetryinterp}, we expect the action on the bimodule types to be still well-defined, although we do not expect there to be an automorphism of the boundary algebra itself. We leave this more general proof of non-contractibility, which will require new technology in the theory of boundary algebras, to future work.

We then generalize the symmetry interpolation method to multiple symmetries. Given $n$ commuting symmetries $U_1,\ldots,U_n$ of a Hamiltonian $H$, we can construct a $T^n$-family via
\begin{equation}\label{eqnTnfamily}
    H(\theta_1,\ldots,\theta_n) \coloneqq U_1(\theta_1) \cdots U_n(\theta_n) H U_n(\theta_n)^\dagger \cdots U_1(\theta_1)^\dagger,
\end{equation}
where $U_j(\theta_j)$ are interpolations as above, which if chosen to satisfy
\begin{equation}
    U_j U_k(\theta_k) = U_k(\theta_k) U_j \qquad \forall j < k,\theta_k,
\end{equation}
then $H(\theta_1,\ldots,\theta_n)$ will be $2\pi$-periodic in each $\theta_k$.

If we choose symmetries which are \emph{fractionalized} on the anyons, meaning that the global commutation relations $U_j U_k = U_k U_j$ locally fail on the anyons \cite{Barkeshli:2014cna}, then we thus obtain families of the second type. In particular, in Section~\ref{sec: 2-para pump II} we will find interpolating generators for a $\mathbb{Z}_2 \times \mathbb{Z}_2$ symmetry which anti-commute on certain anyons. When converted into a $T^2$-family, we obtain a quantized invariant very much analogous to the non-trivial $S^2$-families above.\footnote{Topologically, since the individual 1-cycles of $T^2$ act trivially, we can think of such a $T^2$-family as homotopic to one where we first collapse the usual 1-skeleton of $T^2$ to get $T^2 \to S^2$, and then map $S^2$ around a non-contractible cycle of the moduli space of the toric code phase. Which cycle is determined by the fractionalization pattern.}

From the point of view of the boundary algebra, we can again prove the non-contractibility of these families in the unitary setting.  Exploiting the product structure of the torus, we show that we obtain a ``pump of pumps''. This means that the effective 1+1D boundary $S^1$-family parametrized by $\theta_2$ has a quantized pump invariant which distinguishes $\theta_1 = 0$ and $\theta_1 = 2\pi$.

Finally, by interpolating some of the symmetries and keeping others as global symmetries, we may also obtain symmetry-protected families. In Section~\ref{sec: crystalline pump} we consider interpolating a tensor-product symmetry which fractionalizes with a crystalline rotation symmetry to define a rotation-symmetry protected $S^1$-family.
In this model, the symmetry is no longer internal, but in return we obtain a clear picture in terms of a higher-order abelian anyon pump.

\paragraph{Organization of the paper}
The rest of the paper is organized as follows: 

In Section~\ref{sec: em pump}, we construct $1$-parameter family of toric code systems.
To this end, we focus on the $em$-exchange symmetry inherent in the toric code. 
By interpolating this symmetry, we construct an explicit one-parameter family and observe topological pumping of the $em$-exchange line to the boundary of the system.

In Section~\ref{sec: 2-para pump II}, we construct $2$-parameter families of toric code systems.
To this end, we modefiy the toric code model by introducing additional degrees of freedom on plaquettes, and construct a $\mathbb{Z}_{2}\times\mathbb{Z}_{2}$ symmetry-enriched toric code in which the $\mathbb{Z}_{2}\times\mathbb{Z}_{2}$ symmetry has a nontrivial symmetry fractionalization class.
By interpolating the $\mathbb{Z}_{2}\times\mathbb{Z}_{2}$ symmetry, we construct a $T^{2}$-parameterized model and observe pump of pump phenomena to the boundary of the system.
We also provide Klein bottle families of the toric code phase.

In Section~\ref{sec: crystalline pump}, we construct a model of higher-order anyon pump protected by rotation symmetry.
The model is an $S^{1}$-family of toric codes with $C_{4}$ rotation symmetry, and by varying the parameter, we obtain a pump that transports abelian anyons to the corners of the system.
To show the non-triviality of the family, we perform the gauging of the $1$-form symmetry and confirm that the dual model exhibits a non-trivial crystalline SPT phase.

\section{\texorpdfstring{${\bm 1}$}{1}-parameter family: \texorpdfstring{$\bm{em}$}{em}-exchange defect pumping}
\label{sec: em pump}

In this section, we construct a non-contractible $S^1$-family of Hamiltonians $H(\theta)$ in the toric code phase, characterized by pumping of the $em$-exchange defect, in a way we shall make precise. 
This family will be constructed by symmetry interpolation using a finite depth quantum circuit $U_{em}$ implementing the $em$-exchange symmetry of the toric code Hamiltonian.
To show the non-contractibility of the family, we use the boundary algebra method developed in \cite{jones2023localtopologicalorderboundary} and show that the family induces a non-trivial automorphism of the boundary algebra.

\subsection{Finite depth \texorpdfstring{$em$}{em}-exchange symmetry in the toric code}
\label{sec: em symmetry}

\begin{figure}[t]
\begin{align*}
    \adjincludegraphics[scale=1.25,trim={10pt 10pt 10pt 10pt},valign = c]{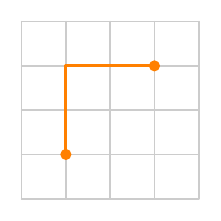}\;
    &\coloneqq
    \adjincludegraphics[scale=1.25,trim={10pt 10pt 10pt 10pt},valign = c]{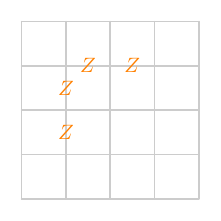},\;\;\;
    \adjincludegraphics[scale=1.25,trim={10pt 10pt 10pt 10pt},valign = c]{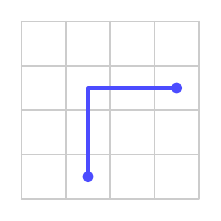}\;
    \coloneqq
    \adjincludegraphics[scale=1.25,trim={10pt 10pt 10pt 10pt},valign = c]{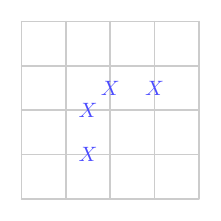}
\end{align*}

    \caption{$e$- and $m$-anyons in the toric code ground state. The $e$-anyon is realized at the endpoints of an $e$-anyon string indicated by orange dots on the vertex. Similarly, the $m$-anyon is realized at the endpoints of an $m$-anyon string indicated by blue dots on the plaquettes.
    }
    \label{fig: anyons}
  \end{figure}

The toric code model is defined on a square lattice with a local Hilbert space $\mathbb{C}^{2}$ on each edge.
Let us consider the following stabilizers:
\begin{equation}
    A_{v} \coloneqq  X_{v_{1}}X_{v_{2}}X_{v_{3}}X_{v_{4}}=
        \vcenter{
        \hbox{
            \adjincludegraphics[scale=1.2,trim={10pt 10pt 10pt 10pt},valign = c]{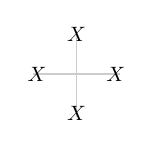}
        }
        },\;\;
    B_{p} \coloneqq  Z_{p_{1}}Z_{p_{2}}Z_{p_{3}}Z_{p_{4}}=
        \vcenter{
        \hbox{
            \adjincludegraphics[scale=1.2,trim={10pt 10pt 10pt 10pt},valign = c]{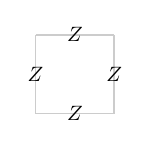}
        }
        },
\end{equation}
where $v_{1},v_{2},v_{3},v_{4}$ are label of the top, right, bottom, and left edges of the vertex $v$, and $p_{1},p_{2},p_{3},p_{4}$ are label of the top, right, bottom, and left edges of the plaquette $p$, and $X,Z$ denote the Pauli $X,Z$ matrices.
The Hamiltonian is given by
\begin{equation}
    H_{TC} \coloneqq  -\sum_{v} A_{v} -\sum_{p} B_{p}.
\end{equation}
An important feature of this model is that it hosts anyonic excitations.  
We first define the $e$-anyon loop operator, which is given by the product of $Z$ along a path $\gamma_e$ on the links:
\begin{equation}
    W_e(\gamma_e) = \prod_{l \in \gamma_e} Z_l.
\end{equation}
Taking $\gamma_e$ to be a closed loop, this operator commutes with the Hamiltonian and therefore does not change the energy.
However, if we consider a path connecting two points instead of a closed loop, it anticommutes with the $A_v$ stabilizers at the two endpoints.
We refer to the quasiparticle localized at such a vertex as an $e$-anyon. See Fig.~\ref{fig: anyons}.

Next, we define the $m$-anyon loop operator, which is given by the product of $X$ operators along a path $\gamma_m$ on the dual lattice:
\begin{equation}
    W_m(\gamma_m) = \prod_{l \cap \gamma_m \neq \varnothing} X_l.
\end{equation}
This operator also commutes with the Hamiltonian if $\gamma_m$ is taken to be a closed loop and thus does not change the energy. 
However, if we consider a path connecting two plaquettes instead of a closed loop, it anticommutes with the $B_p$ stabilizers at the two ends.
We refer to the quasiparticle localized at such a plaquette as an $m$-anyon. See Fig.~\ref{fig: anyons}.
Finally, an anyon consisting of both an $e$-anyon and an $m$-anyon at the same location is called an $f$-anyon. These anyons exhaust the non-trivial superselection sectors of the toric code \cite{Kitaev_2003}.

This Hamiltonian has a symmetry which exchanges the $e$ and $m$ anyons, called the $em$-exchange symmetry. This is usually constructed (see e.g.~\cite{Dennis_2002}) by applying a Hadamard gate
\begin{equation}
    H\coloneq \frac{1}{\sqrt{2}}(X + Z) 
    =  \frac{1}{\sqrt{2}}
    \begin{pmatrix}
        1 & 1\\
        1 & -1
    \end{pmatrix}
\end{equation}
to each site, which transforms
\begin{equation}
    A_v \mapsto \adjincludegraphics[scale=1.2,trim={10pt 10pt 10pt 10pt},valign = c]{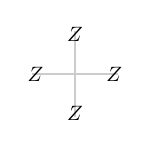},
  \qquad B_p \mapsto \adjincludegraphics[scale=1.2,trim={10pt 10pt 10pt 10pt},valign = c]{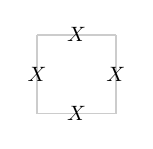}.
\end{equation}
These are the toric code terms on the \emph{dual} lattice. Thus, we must combine such a transformation with a non-trivial motion of the lattice to its dual, such as a $\frac12(1,1)$ translation or a $\pi/2$ rotation about an edge center to obtain a symmetry of the original Hamiltonian.

Since it exchanges vertex and plaquette terms, this symmetry exchanges $e$ and $m$ anyons. However, these are not finite depth quantum circuits, so we cannot interpolate them.\footnote{With the addition of ancillas, one could combine the $\frac12(1,1)$ translation of the toric code lattice with a $-\frac12(1,1)$ translation of the ancilla lattice, which would then be realizable as a circuit. However, our goal is to define a family in the toric code Hilbert space.} To create a parametrized family, we want a different construction of the $em$-exchange symmetry.

For the construction, we will use the Hadamard gate above as well as the CNOT gate:
\begin{equation}
    \mathrm{CNOT}_{i,j}\coloneqq e^{i\frac{\pi}{4}(1-Z_{i})(1-X_{j})}=\adjincludegraphics[scale=1,trim={10pt 20pt 10pt 20pt},valign = c]{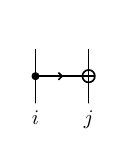}
    % ,\;
    % H \coloneqq  \frac{1}{\sqrt{2}}(X + Z) 
    % =  \frac{1}{\sqrt{2}}
    % \begin{pmatrix}
    %     1 & 1\\
    %     1 & -1
    % \end{pmatrix}.
\end{equation}\label{eq: def CNOT Hadamard}
By using these gates, we can define the $em$-exchange symmetry operator as the following unitary circuit:
\begin{equation}
    U_{em} = \prod_{l} H_{l}\prod_{p} \mathrm{CNOT}_{p_{3},p_{4}}\prod_{p} \mathrm{CNOT}_{p_{2},p_{3}}\prod_{p} \mathrm{CNOT}_{p_{1},p_{2}},
\end{equation}
where, for the plaquette labeled $p$, $p_1$, $p_2$, $p_3$ and $p_4$ label the top, right, bottom, and left sites of the plaquette, respectively.
Diagramatically, the action of this circuit on each plaquette can be represented as follows:
\begin{equation}
    \adjincludegraphics[scale=1.5,trim={10pt 10pt 10pt 10pt},valign = c]{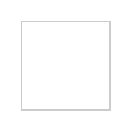}\;
    \to
    \adjincludegraphics[scale=1.5,trim={10pt 10pt 10pt 10pt},valign = c]{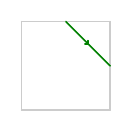}\;
    \to
    \adjincludegraphics[scale=1.5,trim={10pt 10pt 10pt 10pt},valign = c]{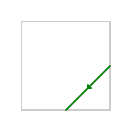}\;
    \to
    \adjincludegraphics[scale=1.5,trim={10pt 10pt 10pt 10pt},valign = c]{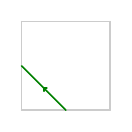}\;
    \to
    \adjincludegraphics[scale=1.5,trim={10pt 10pt 10pt 10pt},valign = c]{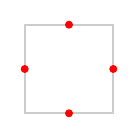}\;
\end{equation}
Here, green lines represent the CNOT gate, with the arrow pointing from $i \to j$ as in \eqref{eq: def CNOT Hadamard}, and red dots represent the Hadamard gate.

\begin{figure}[t]
    \adjincludegraphics[scale=1.0,trim={25pt 10pt 12pt 10pt},valign = c]{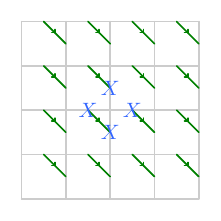}$\;
    \to$
    \adjincludegraphics[scale=1.0,trim={12pt 10pt 12pt 10pt},valign = c]{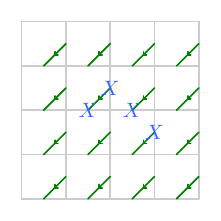}$\;
    \to$
    \adjincludegraphics[scale=1.0,trim={12pt 10pt 12pt 10pt},valign = c]{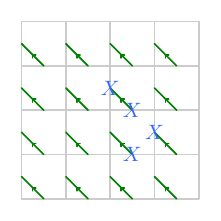}$\;
    \to$
    \adjincludegraphics[scale=1.0,trim={12pt 10pt 12pt 10pt},valign = c]{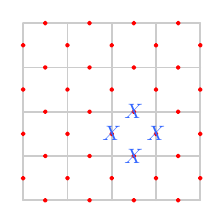}$\;
    \to$
    \adjincludegraphics[scale=1.0,trim={12pt 10pt 15pt 10pt},valign = c]{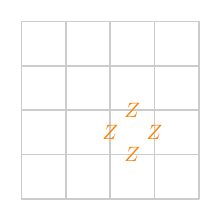}$\;$

    \caption{$U_{em}$-action on the plaquette operator $A_{v}$\\
    }
    \label{fig: em on v}
  \end{figure}

\begin{figure}[t]
    \adjincludegraphics[scale=1.0,trim={25pt 10pt 12pt 10pt},valign = c]{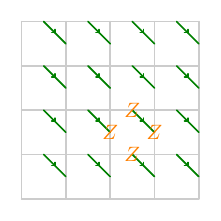}$\;
    \to$
    \adjincludegraphics[scale=1.0,trim={12pt 10pt 12pt 10pt},valign = c]{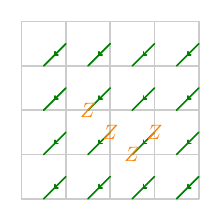}$\;
    \to$
    \adjincludegraphics[scale=1.0,trim={12pt 10pt 12pt 10pt},valign = c]{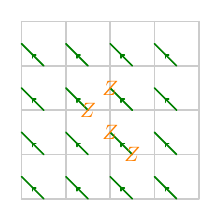}$\;
    \to$
    \adjincludegraphics[scale=1.0,trim={12pt 10pt 12pt 10pt},valign = c]{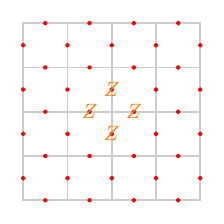}$\;
    \to$
    \adjincludegraphics[scale=1.0,trim={12pt 10pt 15pt 10pt},valign = c]{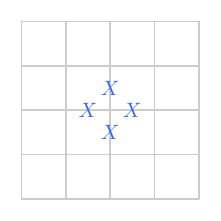}$\;$

    \caption{$U_{em}$-action on the plaquette operator $B_{p}$\\
    }
    \label{fig: em on p}
  \end{figure}

We can show that the above circuit is a symmetry operator of the toric code model.
To see this, the following commutation relations are useful:
\begin{align}
    \adjincludegraphics[scale=1.25,trim={10pt 10pt 10pt 10pt},valign = c]{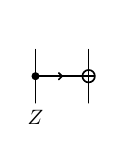}&=
    \adjincludegraphics[scale=1.25,trim={10pt 10pt 10pt 10pt},valign = c]{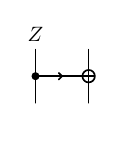},
    \adjincludegraphics[scale=1.25,trim={10pt 10pt 10pt 10pt},valign = c]{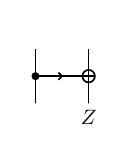}=
    \adjincludegraphics[scale=1.25,trim={10pt 10pt 10pt 10pt},valign = c]{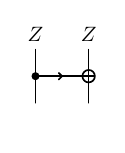},
    \adjincludegraphics[scale=1.25,trim={10pt 10pt 10pt 10pt},valign = c]{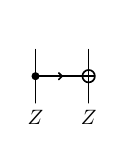}=
    \adjincludegraphics[scale=1.25,trim={10pt 10pt 10pt 10pt},valign = c]{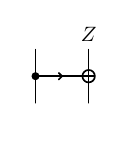},\\
    \adjincludegraphics[scale=1.25,trim={10pt 10pt 10pt 10pt},valign = c]{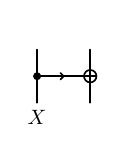}&=
    \adjincludegraphics[scale=1.25,trim={10pt 10pt 10pt 10pt},valign = c]{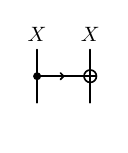},
    \adjincludegraphics[scale=1.25,trim={10pt 10pt 10pt 10pt},valign = c]{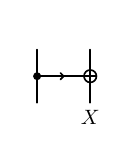}=
    \adjincludegraphics[scale=1.25,trim={10pt 10pt 10pt 10pt},valign = c]{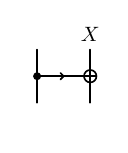},
    \adjincludegraphics[scale=1.25,trim={10pt 10pt 10pt 10pt},valign = c]{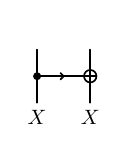}=
    \adjincludegraphics[scale=1.25,trim={10pt 10pt 10pt 10pt},valign = c]{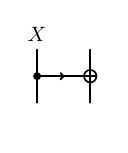}.
\end{align}
This leads to the following action of $U_{em}$ on single Pauli operators:
\begin{align}\label{eqnsinglepauliaction}
    &\;\nonumber\\
    \adjincludegraphics[scale=1.25,trim={10pt 25pt 10pt 25pt},valign = c]{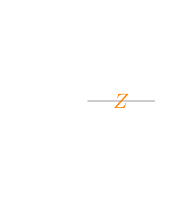}
    &\mapsto
    \adjincludegraphics[scale=1.25,trim={10pt 25pt 10pt 25pt},valign = c]{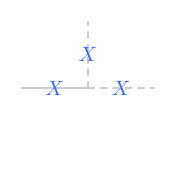},
    \adjincludegraphics[scale=1.25,trim={10pt 25pt 20pt 25pt},valign = c]{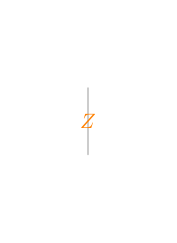}
    \mapsto
    \adjincludegraphics[scale=1.25,trim={10pt 25pt 10pt 25pt},valign = c]{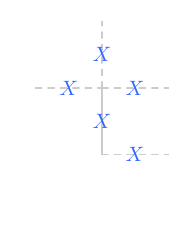},\\
    \adjincludegraphics[scale=1.25,trim={10pt 35pt 10pt 35pt},valign = c]{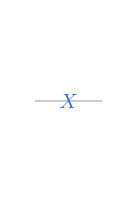}
    &\mapsto
    \adjincludegraphics[scale=1.25,trim={10pt 35pt 10pt 35pt},valign = c]{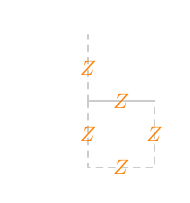},
    \adjincludegraphics[scale=1.25,trim={10pt 35pt 20pt 35pt},valign = c]{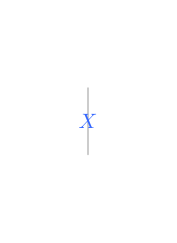}
    \mapsto
    \adjincludegraphics[scale=1.25,trim={10pt 35pt 10pt 35pt},valign = c]{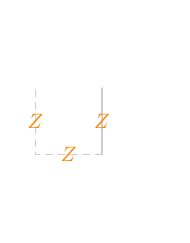},\\
    \;\nonumber
\end{align}
Here the solid link is the same on the left and right sides, while other links are drawn with dashed lines. Using these rules, it can be verified that $U_{em}$ exchanges $A_{v}$ and $B_{p}$ terms. See Fig.~\ref{fig: em on v} and Fig.~\ref{fig: em on p}. Therefore, $U_{em}$ is a symmetry of the toric code with periodic boundary conditions or in the infinite plane. The same circuit symmetry was recently discussed in \cite{Tu:2025bqf,Shirley:2025yji}.

We also see from the above that $U_{em}$ maps closed $e$-strings to closed $m$-strings and vice versa, since these are locally products of $A_v$ and $B_p$ terms. These closed strings detect the $e$ and $m$ superselection sectors by braiding \cite{Kitaev_2003}, so therefore $U_{em}$ exchanges anyons of $e$ and $m$ type.

Note that this construction has $U_{em}^2 \neq 1$ (in fact $U_{em}^4 = 1$, as can be checked from \eqref{eqnsinglepauliaction}), while the action on the braided fusion category is a $\mathbb{Z}_2$ action. The lack of a $\mathbb{Z}_2$ group law on the lattice does not pose an obstacle for defining a periodic family via interpolation. A $\mathbb{Z}_2$ realization may be found in \cite{Tu:2025bqf}.


\subsection{Partial symmetry action and symmetry defect}
\label{sec: sym fractionalization}

When $U_{em}$ is applied to the whole system, it preserves the ground state. 
However, when applied partially in a region such as the right half-plane, it creates a one dimensional ``symmetry defect'' localized near the boundary of the region where it was applied. When we interpolate the symmetry to a $1$-parameter periodic family as in \eqref{eqnsymmetryinterp}, we can think of this defect being pumped to the boundary around each cycle.

\begin{figure}[t]
\begin{align*}
    \adjincludegraphics[scale=1.25,trim={10pt 10pt 10pt 10pt},valign = c]{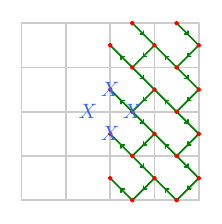}\;
    &\to
    \adjincludegraphics[scale=1.25,trim={10pt 10pt 10pt 10pt},valign = c]{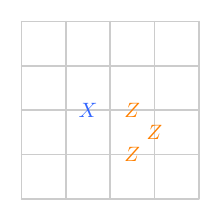},\;\;\;
    \adjincludegraphics[scale=1.25,trim={10pt 10pt 10pt 10pt},valign = c]{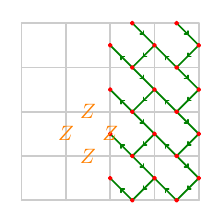}\;
    \to
    \adjincludegraphics[scale=1.25,trim={10pt 10pt 10pt 10pt},valign = c]{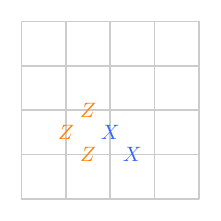}\;\\
    \adjincludegraphics[scale=1.25,trim={10pt 10pt 10pt 10pt},valign = c]{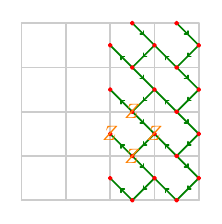}\;
    &\to
    \adjincludegraphics[scale=1.25,trim={10pt 10pt 10pt 10pt},valign = c]{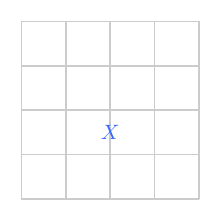}\;
    % \adjincludegraphics[scale=1,trim={10pt 10pt 10pt 10pt},valign = c]{figs/lattice_defect.pdf}\;
\end{align*}

    \caption{partial symmetry action on the stabilizers\\
    }
    \label{fig: frac sym on stab}
  \end{figure}

Let us describe the symmetry defect obtained from a partial application of our circuit. We define $U_{em}^{R}$ as the restriction of $U_{em}$ to the right half-plane shown in Fig.~\ref{fig: frac sym on stab}. This will preserve stabilizers $A_v$ and $B_p$ supported a certain distance from the boundary.

We observe that the plaquette operators on the right side of the boundary are mapped to the $X$ operators on the boundary. 
As a result, the spins on the boundary are completely fixed by these stabilizers. 
Meanwhile, the plaquette operators on the left side and the vertex term on the boundary are mapped to a combination of Pauli Z and X, as shown in Fig.~\ref{fig: frac sym on stab}.
The resulting stabilizers are as follows:
\begin{align}
    \adjincludegraphics[scale=1.25,trim={10pt 10pt 10pt 10pt},valign = c]{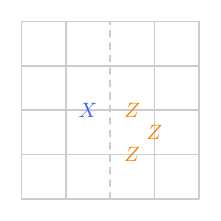}\;\;\;\;\;\;\;\;\;\;
    \adjincludegraphics[scale=1.25,trim={10pt 10pt 10pt 10pt},valign = c]{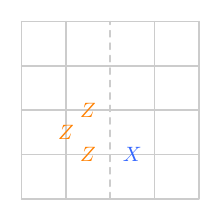}
\end{align}
where the spins on the dashed line are missing. The two sides of the cut can be considered as rough boundaries with these extra two stabilizers. This defect can be related to the well-known $em$-exchange defect \cite{Kitaev:2011dxc,Bombin_2010,Bridgeman:2017gyy,Barkeshli:2022wuz,Tantivasadakarn:2023zov, Vanhove:2024lmj, li2024domainwallssptsewing} by shifting one half of the lattice:
\begin{equation}\label{eq: defect stab}
    \adjincludegraphics[scale=1.25,trim={10pt 10pt 10pt 10pt},valign = c]{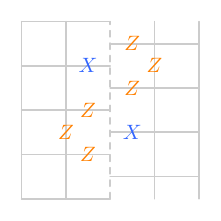}
\end{equation}

\begin{figure}[ht]
\begin{align*}
    \adjincludegraphics[scale=1.25,trim={10pt 10pt 10pt 10pt},valign = c]{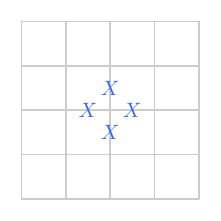}\;
    &\to
    \adjincludegraphics[scale=1.25,trim={10pt 10pt 10pt 10pt},valign = c]{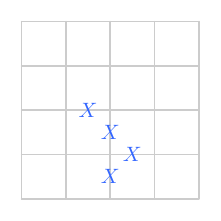},\;\;\;
    \adjincludegraphics[scale=1.25,trim={10pt 10pt 10pt 10pt},valign = c]{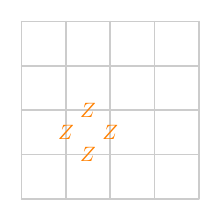}\;
    \to
    \adjincludegraphics[scale=1.25,trim={10pt 10pt 10pt 10pt},valign = c]{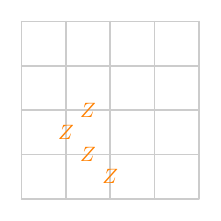}\;\\
    \adjincludegraphics[scale=1.25,trim={10pt 10pt 10pt 10pt},valign = c]{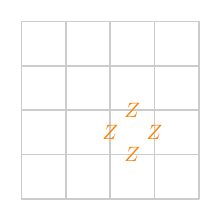}\;
    &\to
    \adjincludegraphics[scale=1.25,trim={10pt 10pt 10pt 10pt},valign = c]{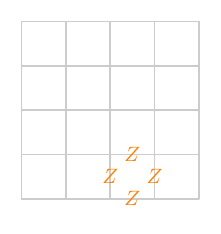}\;
\end{align*}
    \caption{
        The action of $(U_{em}^{R})^4$. By applying $(U_{em}^{R})^4$, a one-site translation is realized.
    }
    \label{fig: frac sym on stab four times}
  \end{figure}

Note that, as shown in Fig.~\ref{fig: frac sym on stab four times}, applying $U_{em}^{R}$ four times induces a one-site translation. In \cite{Tu:2025bqf} this was used to argue that there is no finite depth circuit $C$ such that $C U_{em}C^\dagger$ is an on-site symmetry.

\begin{figure}[t]
    \begin{minipage}[b]{0.2\columnwidth}
        \centering
        \adjincludegraphics[scale=1.5,trim={20pt 10pt 10pt 10pt},valign = c]{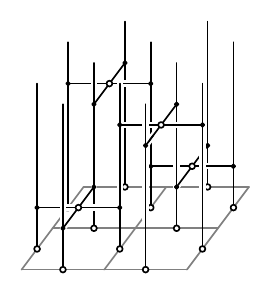}
        \caption*{(a)}
      \end{minipage}
      \hspace{0.04\columnwidth}
      \begin{minipage}[b]{0.75\columnwidth}
        \centering
        \adjincludegraphics[scale=1.5,trim={0pt 10pt 10pt 10pt},valign = c]{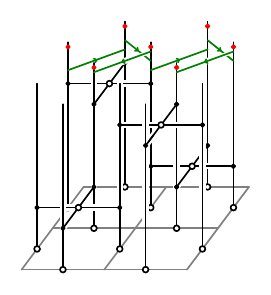}
        =
        \adjincludegraphics[scale=1.5,trim={10pt 10pt 30pt 10pt},valign = c]{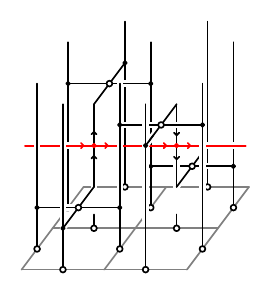}
        \caption*{(b)}
      \end{minipage}
  
    \caption{
    (a) The tensor network representation of the toric code ground state.
    The white circles represent the multiplication of $\mathbb{C}[G]$, and the black circles represent the comultiplication of $\mathbb{C}[G]$.
    (b) A partial symmetry action of $U_{em}$ and its symmetry fractionalization.
    The left-hand side represents the partial symmetry action of $U_{em}$ to the plaquettes on the far side, and the right-hand side represents its symmetry fractionalization.
    }
    \label{fig: tensor network representation}
  \end{figure}

From the above discussion, we found that the state obtained by applying $U_{em}^{R}$ corresponds to the stabilizer state for \eqref{eq: defect stab}. 
By using this characterization, we can see that the fractionalized symmetry acts as a Kramers-Wannier line on the bond Hilbert space in the tensor network representation of the toric code: 
First, the ground state of the toric code can be represented as a tensor network as shown in Fig.~\ref{fig: tensor network representation}~(a). 
When $U_{em}$ is partially applied only to the plaquettes on the far side, the resulting state can be represented as shown in Fig.~\ref{fig: tensor network representation}~(b). 
Here, the Kramers-Wannier line is represented by the red line:
\begin{equation}
    \adjincludegraphics[scale=1.5,trim={10pt 10pt 10pt 10pt},valign = c]{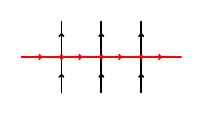}
    \;\;\coloneqq \;\;
    \adjincludegraphics[scale=1.5,trim={10pt 10pt 10pt 10pt},valign = c]{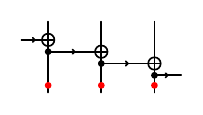}
\end{equation}
where, each red circle in the right-hand side represents the Hadamard gate.
Then, this tensor network state can be verified to be the stabilizer state for \eqref{eq: defect stab}. This Kramers-Wannier action in the bond Hilbert space reflects the action on the boundary algebra we will see shortly.

\subsection{Definition of the \texorpdfstring{$S^{1}$}{S1}-family with \texorpdfstring{$em$}{em}-exchange from interpolation}
\label{sec: 1-para}
In general, for a Hamiltonian $H$ that possesses a finite-depth quantum-circuit symmetry, one can construct an $S^{1}$-family of Hamiltonians as follows:
Suppose the Hamiltonian has a symmetry that is implemented by a finite-depth quantum circuit $U\coloneqq \prod_{i=1}^{N} L_i$. 
Each layer $L_i$ in the circuit consists of local unitary operators whose supports do not overlap, $L_i = \prod_{p} u_p$. 
Since every $u_p$ is a finite-dimensional unitary operator, it can be connected continuously to the identity operator by $u_p(\theta)\coloneqq e^{N\frac{\theta}{2\pi}\log(u_p)}$, and one can define an interpolation of each layer by $L_i(\theta)\coloneqq \prod_p u_p(\theta)$. Then, the interpolation of the symmetry operator is given by
\begin{equation}
    U(\theta)\coloneqq  L_{n+1}(\theta-\frac{2\pi n}{N})\prod_{i=1}^{n} L_i,\qquad \theta\in [\frac{2\pi n}{N},\frac{2\pi (n+1)}{N}],
\end{equation}
for $n\in{0,...,N-1}$, and an $S^1$-family of the Hamiltonian is defined by
\begin{equation}
    H(\theta)\coloneqq U(\theta)H U(\theta)^{\dagger}.
\end{equation}
Note that $U(\theta)$ is not $2\pi$-periodic, whereas $H(\theta)$ is, because $U(2\pi) H U(2\pi)^\dagger = U H U^\dagger = H$.
Moreover, $H(\theta)$ is local since $U(\theta)$ is a finite-depth quantum circuit for every $\theta \in [0,2\pi]$.

Let us apply the procedure to the $em$-exchange symmetry $U_{em}$ of the toric code Hamiltonian. The interpolation of the symmetry $U_{em}$ is defined by
\begin{equation}\label{eqnlayerwiseinterp}
    U_{em}(\theta) = \left\{\begin{aligned}
        \prod_{p} \mathrm{CNOT}(\theta)_{p_{1},p_{2}},& \;\;\theta\in [0,\frac{\pi}{2}],\\
        \prod_{p} \mathrm{CNOT}(\theta)_{p_{2},p_{3}}\prod_{p} \mathrm{CNOT}_{p_{1},p_{2}},&  \;\;\theta\in [\frac{\pi}{2},\pi],\\
        \prod_{p} \mathrm{CNOT}(\theta)_{p_{3},p_{4}}\prod_{p} \mathrm{CNOT}_{p_{2},p_{3}}\prod_{p} \mathrm{CNOT}_{p_{1},p_{2}},&  \;\;\theta\in [\pi,\frac{3\pi}{2}],\\
        \prod_{l} H_{l}(\theta)\prod_{p} \mathrm{CNOT}_{p_{3},p_{4}}\prod_{p} \mathrm{CNOT}_{p_{2},p_{3}}\prod_{p} \mathrm{CNOT}_{p_{1},p_{2}},&  \;\;\theta\in [\frac{3\pi}{2},2\pi],
    \end{aligned}\right.
\end{equation}
for $\theta\in\left[0,2\pi\right]$, where
\begin{equation}
    H_{l}(\theta) = e^{i4\frac{\theta}{4}H_{l}},\;\; \mathrm{CNOT}(\theta)_{p,p'} = e^{i4\frac{\theta}{4}\mathrm{CNOT}_{pp^{\prime}}},
\end{equation}
$H_l$ is the Hadamard gate on the link $l$, and $p_1$,$p_2$,$p_3$ and $p_4$ label the top, right, bottom, and left links of the plaquette $p$, respectively.
The family is defined as follows:
\begin{equation}
    H_{em}(\theta) \coloneqq  U_{em}(\theta) H_{TC} U_{em}(\theta)^{\dagger}.
\end{equation}

\subsection{Boundary Algebras and Nontriviality of the \texorpdfstring{$S^{1}$}{S1}-family}
\label{sec: toric code boundary algebra}

In this section, we show that the $S^{1}$-family constructed in Section~\ref{sec: 1-para} is non-contractible as a unitary family, in the sense explained below \eqref{eqnsymmetryinterp}. In particular, we cannot find an extension of the finite depth unitary $U_{em}(\theta)$ over a disc $D^2$.

The key idea is the definition of a boundary algebra \cite{jones2023localtopologicalorderboundary}. To define the boundary algebra, we need to specify a commuting projector truncation of $H$, $H_b$ to a half space. Intuitively, Let $P_b$ be the projection to the ground state subspace of $H_b$ (which may be extensive in size). The boundary algebra is the algebra of operators of the form $P_b a P_b$, where $a$ is a local operator. Note that if $a_1$ and $a_2$ act identically in the ground state subspace, then $P_ba_1P_b = P_ba_2 P_b$. Assuming that the bulk is topologically ordered (LTO axioms of \cite{jones2023localtopologicalorderboundary}), the algebra of these operators is given by $P_baP_b$ where $a$ is supported in a finite width neighborhood of the boundary, hence the name ``boundary algebra''.

\begin{figure}[t]
    \centering
    \includegraphics[width=0.4\linewidth]{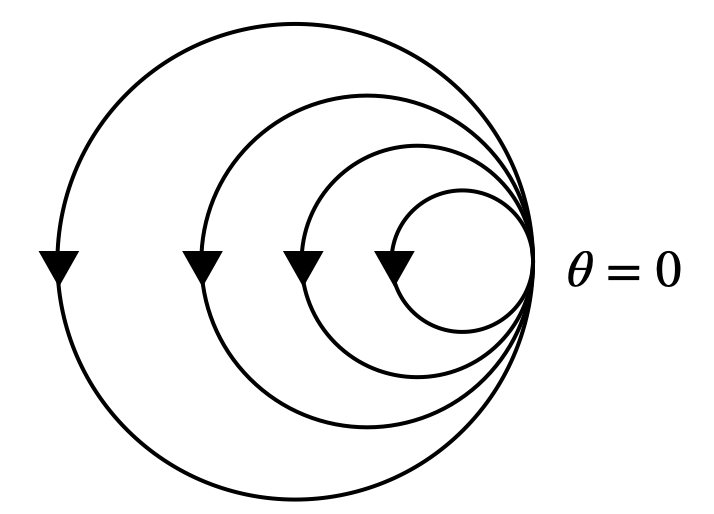}
    \caption{A one parameter family of loops in the unit disc, beginning and ending at the reference point $r = 1$, $\theta = 0$ corresponding to the interpolated boundary family $H_b(\theta)$. The loop around the boundary $r = 1$ produces an automorphism $U_b$ of the boundary algebra. The automorphism associated to the shrinking loops, we obtain boundary automorphism interpolating from $U_b$ to the identity. Thus, if the boundary automorphism can be associated with a deformation-invariant index distinct from the identity, then we have a contradiction, and the family must be non-contractible.}
    \label{fig:deformation-of-loops}
\end{figure}

To extend this definition to families, we need a family of commuting projector Hamiltonians $H_b$. For families obtained by symmetry interpolation, this can be achieved by truncating the symmetry $U$ to be interpolated to the halfspace. This must be done so that the truncated operator $U_b$ is a symmetry of $H_b$. Assuming this, we can form the interpolated family $H_b(\theta) = U_b(\theta) H_b U_b(\theta)^\dagger$. This will be periodic, and in particular we can identify the boundary algebras at $\theta = 0$ and $\theta = 2\pi$. However, they come identified by the automorphism $U_b$, which naturally acts on the boundary algebra via
\begin{equation}
    U_b P_b a P_b U_b^\dagger = P_b U a U^\dagger P_b
\end{equation}
since it is a symmetry. However, although we can interpolate $U_b$ to the identity as a bulk symmetry, we may not be able to do so as an automorphism of the boundary algebra. In this case, we have a non-contractible family, see Fig.~\ref{fig:deformation-of-loops}. We will show this is the case for the toric code family below.

\begin{figure}[t]
    \centering
    \begin{minipage}{0.32\linewidth}
        \adjincludegraphics[scale=1.5,trim={10pt 10pt 10pt 10pt},valign = c]{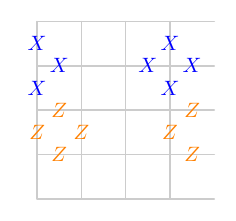}
      \caption*{(a)}
    \end{minipage}
    \hfill
    \begin{minipage}{0.32\linewidth}
        \adjincludegraphics[scale=1.5,trim={10pt 10pt 10pt 10pt},valign = c]{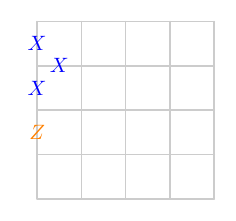}
      \caption*{(b)}
    \end{minipage}
    \hfill
    \begin{minipage}{0.32\linewidth}
        \adjincludegraphics[scale=1.5,trim={10pt 10pt 10pt 10pt},valign = c]{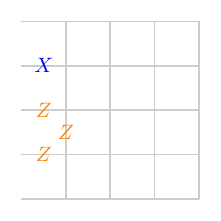}
      \caption*{(c)}
    \end{minipage}
  
    \caption{
        The boundary types and boundary algebras of the toric code.
        (a) A smooth boundary (left side) and a rough boundary (right side).
        On the smooth boundary, the stabilizer $A_{v}$ is replaced with a stabilizer that applies $X$ to the two boundary links and to the adjacent interior link.
        On the rough boundary, the stabilizer $B_{p}$ is replaced with a stabilizer that applies $Z$ to the two boundary links and to the adjacent interior link.
        (b) The generators of the boundary algebra for the smooth boundary.
        (c) The generators of the boundary algebra for the rough boundary. 
    }
    \label{fig: toric_bdy}
\end{figure}

\paragraph{Boundary types and boundary algebras} Usually when one discusses toric code boundaries one has in mind either a ``smooth'' or ``rough'' boundary, shown on the left and right sides of Fig.~\ref{fig: toric_bdy} (a). Here one typically includes in $H_b$ all the bulk $A_v$ and $B_p$ terms which fit in the truncated lattice as well as some extra terms such as the 3-$X$ term on the left (smooth) side of Fig.~\ref{fig: toric_bdy} (a) and the 3-$Z$ term on the right (rough) side of Fig.~\ref{fig: toric_bdy} (a) \cite{Bravyi:1998sy,Kitaev_2006,Kitaev:2011dxc}. These are ``topological boundaries'', for which the boundary algebra, which consists of local operators preserving the ground state space, are trivial. We do not want to consider these.

Instead, to define a non-trivial boundary algebra, we want only to keep the bulk $A_v$ and $B_p$ terms which fit letting these define $H_b$, and consider all of the operators which commute with these terms. The generators of the boundary algebras for either a smooth or rough edge are shown in Fig.~\ref{fig: toric_bdy} (b) and Fig.~\ref{fig: toric_bdy} (c), respectively.\footnote{See also \cite{Chatterjee:2022kxb,Inamura:2023ldn} for a related discussion where these are called patch operators.}

More precisely, in the case of the smooth boundary in Fig.~\ref{fig: toric_bdy} (b), the operator $S_{v}$ that applies three $X$ to the three links adjacent to a boundary vertex $v$ commutes with all bulk stabilizers:
\begin{equation}
    S_{v} \coloneq X_{l_{1}}X_{l_{2}}X_{l_{3}},
\end{equation}
where $l_{1}$ and $l_{2}$ are the two boundary links adjacent to the vertex $v$ and $l_{3}$ is the adjacent interior link.
Also, an operator $S_{l}$ that applies $Z$ to a link $l$ on the boundary commutes with all bulk stabilizers (see Fig.~\ref{fig: toric_bdy}~(b)):
\begin{equation}
    S_{l} \coloneq Z_{l}.
\end{equation}
Thus, these operators are elements of the boundary algebra for the smooth boundary , and moreover it is known that they generate the boundary algebra, in the sense that all elements of the boundary algebra are sums of products of $P_b S_v P_b$ and $P_b S_l P_b$ for different $v$ and $l$ \cite{jones2023localtopologicalorderboundary}.

Similarly, for a rough boundary, operators $R_{p}$ that apply three $Z$ to the links adjacent to a boundary plaquette $p$ and operators $R_{l}$ that apply $X$ to a link $l$ are elements of the boundary algebra for the rough boundary, and they generate the boundary algebra (see Fig.~\ref{fig: toric_bdy}~(c)):
\begin{equation}
    R_{p} \coloneq Z_{l_{1}}Z_{l_{2}}Z_{l_{3}},\qquad
    R_{l} \coloneq X_{l},
\end{equation}
where $l_{1}$ and $l_{2}$ are the two boundary links adjacent to the plaquette $p$ and $l_{3}$ is the adjacent interior link.

Both of these algebras are in fact isomorphic.\footnote{The right notion of isomorphism is a bounded-spread isomorphism, which is quite analogous to a QCA. See \cite{jones2023localtopologicalorderboundary} for precise definitions.} For the smooth boundary, consider a 1d spin-$1/2$ chain associated with the parallel boundary edges. Let $l_1,l_2$ be the boundary edges meeting the vertex $v$. Consider the map
\[S_v \mapsto \tilde Z_{l_1} \tilde Z_{l_2} \\
S_l \mapsto \tilde X_l.\]
On a half-infinite plane, this gives an injective algebra map from the smooth boundary algebra into the algebra of local operators in the spin-$1/2$ chain. Its image is the algebra of $\mathbb{Z}_2$-symmetric operators, i.e. those operators commuting with $\prod_l \tilde X_l$. Likewise for the rough boundary, we may consider a spin-$1/2$ chain associated to the perpendicular edges, and define
\[R_p \mapsto \tilde Z_{l_1} \tilde Z_{l_2} \\
R_l \mapsto \tilde X_l,\]
where $l_1,l_2$ are the neighboring boundary edges of a boundary plaquette. This defines an injective algebra map onto the same $\mathbb{Z}_2$-symmetric subalgebra of the spin-$1/2$ chain.

\paragraph{Automorphism of boundary algebras induced by \texorpdfstring{$U_{em}$}{Uem}}
Let us consider the system with a boundary to demonstrate the nontriviality of the constructed $S^{1}$-family $H(t)$.

\begin{figure}[t]
    \centering
    \begin{minipage}{0.28\linewidth}
        \adjincludegraphics[scale=1.5,trim={10pt 10pt 10pt 10pt},valign = c]{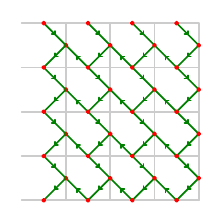}
      \caption*{(a)}
    \end{minipage}
    \hfill
    \begin{minipage}{0.63\linewidth}
        \adjincludegraphics[scale=1.5,trim={10pt 10pt 10pt 10pt},valign = c]{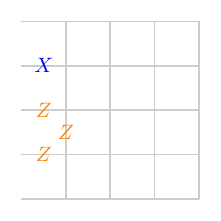}
        $\;\;\longrightarrow\;\;$
        \adjincludegraphics[scale=1.5,trim={10pt 10pt 10pt 10pt},valign = c]{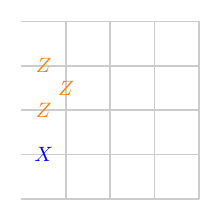}
      \caption*{(b)}
    \end{minipage}
  
    \caption{
        The unitary circuit truncated near the boundary and the automorphism of the boundary algebra induced by \texorpdfstring{$U_{em}$}{Uem}.
        (a) The unitary circuit truncated near the boundary. We get rid of the CNOT and Hadamard gates whose supports are not entirely contained in the semi-infinite plane.
        (b) The automorphism of the boundary algebra induced by $U_{em}$.
        The left-hand side represents the generators of the boundary algebra for the rough boundary, and the right-hand side represents their images under the automorphism.
    }
    \label{fig: toric_bdy_auto}
\end{figure}

We consider the rough boundary.
To define the family with a boundary, we must specify how to cut $U_{em}(t)$ near the boundary.
The unitary $U_{em}$ was defined as a product of CNOT and Hadamard gates.
If we take only those CNOT and Hadamard gates whose entire supports lie within the semi-infinite plane (see Fig.~\ref{fig: toric_bdy_auto}~(a)), this turns out to be a symmetry of $H_b$, as can be seen from the action on the $A_v$ and $B_p$ terms in Fig.~\ref{fig: em on v} and Fig.~\ref{fig: em on p}. Let this define $U_{em}^{bdy.}$ and let $U_{em}^{bdy.}(t)$ be the layerwise interpolation as in \eqref{eqnlayerwiseinterp}.

Then, by varying $t$ from $0$ to $2\pi$, the generators of the boundary algebra are transformed as follows (see Fig.~\ref{fig: toric_bdy_auto}~(b)):
\begin{equation}
    R_{p}\mapsto R_{l},\qquad
    R_{l}\mapsto R_{p}.
\end{equation}
Therefore, as operators on the space of boundary conditions, they are transformed as
\begin{equation}
    \widetilde{Z}_{l_{1}}\widetilde{Z}_{l_{2}}\mapsto \widetilde{X}_{l_{2}},\qquad
    \widetilde{X}_{l_{1}}\mapsto \widetilde{Z}_{l_{1}}\widetilde{Z}_{l_{2}}.
\end{equation}
This is nothing but the Kramers-Wannier duality transformation on the effective spins. In \cite{Jones:2023imy,Jones_2026} it was described how to define a quantized invariant for automorphisms like the above in this algebra, which is a generalization of the GNVW invariant for translations in a 1d tensor product algebra. This invariant is valued in integers and is non-trivial for the Kramers-Wannier transformation above, so the family is non-contractible.

\paragraph{Automorphism of DHR bimodules induced by \texorpdfstring{$U_{em}$}{Uem}} It was also demonstrated in \cite{Jones_2026} that the Kramers-Wannier transformation permutes the DHR bimodules corresponding to the $e$ and $m$ anyon excitations. We provide a simple account of this as follows.

Consider a lattice on $[0,L] \times \mathbb{R}$ of finite width $L$ along the $x$ axis and infinite in the $y$ axis, with smooth boundaries at $x = 0$ and $x = L$. Choose $L$ to be several sites, enough so that the two boundary algebras are non-overlapping. We can consider the Hamiltonian $H_s$ of $A_v$ and $B_p$ terms and the associated projector $P_s$.

\begin{figure}[t]
    \centering
    \begin{minipage}{0.23\linewidth}
        \adjincludegraphics[scale=1.1,trim={10pt 10pt 10pt 10pt},valign = c]{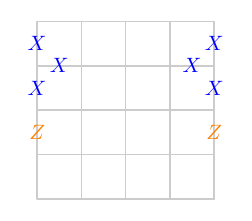}
      \caption*{(a)}
    \end{minipage}
    \hfill
    \begin{minipage}{0.23\linewidth}
        \adjincludegraphics[scale=1.1,trim={10pt 10pt 10pt 10pt},valign = c]{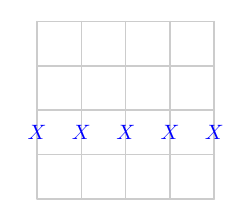}
      \caption*{(b)}
    \end{minipage}
    \hfill
    \begin{minipage}{0.23\linewidth}
        \adjincludegraphics[scale=1.1,trim={10pt 10pt 10pt 10pt},valign = c]{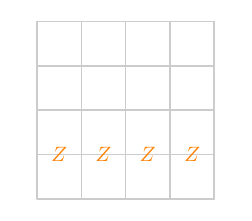}
      \caption*{(c)}
    \end{minipage}
    \hfill
    \begin{minipage}{0.23\linewidth}
        \adjincludegraphics[scale=1.1,trim={10pt 10pt 10pt 10pt},valign = c]{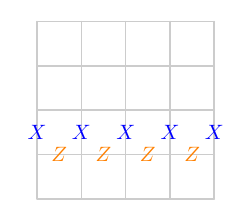}
      \caption*{(d)}
    \end{minipage}

    \caption{A slab algebra defined by two smooth boundaries on the left and right. In addition to the boundary algebras supported on the two edges (we draw two local generators each of the left and right boundary algebras in (a)), there are anyon string operators stretching from one edge to the other. Taking one such string operator in each super selection sector ((b), (c), (d) correspond to the three nontrivial superselection sectors), we get a module basis for the slab algebra. This decomposition is the decomposition into the indecomposable DHR bimodules.}
    \label{fig-DHR-string}
\end{figure}
We define the ``slab algebra'' $\mathcal{A}_s$ to be the operators of the form $P_s a P_s$ where $a$ is a local operator. Let $\mathcal{A}_L$ and $\mathcal{A}_R$ be the left and right boundary algebras. $\mathcal{A}_s$ contains $\mathcal{A}_L \otimes \mathcal{A}_R$ and may be considered a module over it. In fact it is a finitely-generated module, with generators given by anyon string operators stretching from one boundary to the other. See Fig.~\ref{fig-DHR-string}. Each superselection sector corresponds to an indecomposable submodule with is precisely the indecomposable DHR bimodule \cite{Jones_2026}. Since the bulk transformation $U_{em}$ exchanges the types of anyon strings, the induced automorphism of the boundary algebra thus permutes the DHR bimodules according to the same anyon permutation.

One can also consider the smooth boundary.
As in the rough boundary case, we define the boundary-truncated circuit $U_{em}^{s}(t)$ by taking only those CNOT and Hadamard gates whose entire supports lie within the semi-infinite plane.
This is essentially the same computation as in the partial symmetry action in Section~\ref{sec: sym fractionalization}, and we find that the generators of the boundary algebra are transformed as follows:
\begin{equation}
    S_{v}\mapsto R_{p},\qquad
    S_{l}\mapsto R_{l}.
\end{equation}
This again corresponds to a Kramers-Wannier transformation.

\section{\texorpdfstring{${\bm 2}$}{2}-parameter family : pump of \texorpdfstring{${\bm S^1}$}{S1}-family}
\label{sec: 2-para pump II}
In this section, we provide a $\mathbb{Z}_2\times\mathbb{Z}_2$ $0$-form symmetry with a nontrivial symmetry fractionalization class \cite{Barkeshli:2014cna} in the toric code model by introducing additional degrees of freedom. Then by using the symmetry, we construct nontrivial $2$-parameter family models of the toric code phase. To demonstrate the non-triviality of the family, we see the pumping of a nontrivial $S^1$-parameter family along changing one of two parameters. The idea of such a flow of lower-dimensional families are also considered in \cite{Hsin:2022iug,Wen:2021gwc}.

In Section~\ref{sec: em pump}, we constructed an $S^1$-parameter family by interpolating the $\mathbb{Z}_2$ $em$-exchange $0$-form symmetry in the toric code. Since intrinsic $2$-parameter families correspond to codimension-$2$ invertible symmetry, i.e., $1$-form symmetry, one may expect that we can construct a $2$-parameter family model by interpolating a $1$-form symmetry of the toric code. However, such a procedure is unclear because $1$-form symmetries act as line operators in $(2+1)$-dimensional theories. 
To overcome the difficulty, we realize a $\mathbb{Z}_2\times\mathbb{Z}_2$ $0$-form symmetry with a nontrivial symmetry fractionalization class, and construct a $2$-parameter model by interpolating the symmetry. 
This symmetry action traps an anyon at the junction of the two $\mathbb{Z}_2$ symmetry defects, so the corresponding family traps the anyon in the textured picture.

\subsection{\texorpdfstring{$G$}{G}-symmetry enriched toric code}

\paragraph{Symmetry Fractionalization in \texorpdfstring{$G$}{G}-SET}
We briefly recap the general theory developed in \cite{ENO2010,Barkeshli:2014cna}. Consider a $G$ symmetry of a topologically ordered state whose anyonic statistics is described by a modular tensor category $\mathcal{B}$. This symmetry defines a permutation action $\rho$ of $G$ on the superselection sectors of anyons (simple objects in $\mathcal{B}$). Symmetry fractionalization occurs when the non-trivial superselection sectors carry fractional charges under the $G$ action. These are classified by a torsor over $\mathrm{H}^2_{\rho}(BG,\mathcal{A})$, where $\mathcal{A}$ is the group of abelian anyons. When $\rho$ is trivial, the cohomology is just usual group cohomology $\mathrm{H}^2(BG,\mathcal{A})$, which classifies extensions of $G$ by $\mathcal{A}$.

An operational meaning of symmetry fractionalization can be obtained by thinking about
truncating the $G$ action to a region $M$. When $M$ has no boundary (such as on the plane or a torus), these form a genuine unitary representation $U(\cdot)$ when acting on spatial manifolds without boundary. Nevertheless, for $M$ manifolds with boundaries, the representation does not necessarily hold on the ground state space $\ket{\text{GS}}$ and instead obey
\begin{equation}\label{eq:sym_frac_gs}
    U_{M}(g)U_M(h)\left(U_{M}(gh)\right)^{-1}\ket{\text{GS}}=\eta_{\partial M}(g,h)\ket{\text{GS}}.
\end{equation}
Since the left-hand side of this equation acts identically inside $M$, $\eta_{\partial M}(g,h)$ is supported only in the vicinity of $\partial M$. The possible non-triviality of this boundary term is what leads to symmetry fractionalization.  

Following the theme of the previous section, this can be made precise if we choose a truncation $U_M(g)$ acting on the boundary algebra along $\partial M$. As automorphisms of the boundary algebra, $U_M(g)$ will satisfy the group law, but acting on operators in a non-trivial superselection sector, such as the strings stretching between the two boundaries of the slab in Fig.~\ref{fig-DHR-string}, the action on the left (or right) boundary algebra may only hold up to some phase factors multiplying the strings. These superselection-sector-dependent phase factors can be identified with abelian anyons via their braiding phases, and thus we obtain a 2-cocycle defining the symmetry fractionalization class in ${\rm H}^2(BG,\mathcal{A})$.

\paragraph{Simple model of $\mathbb{Z}_{2}\times\mathbb{Z}_{2}$-symmetry enriched toric code}
Let us consider the case where the topologically ordered system is the toric code and $G=\mathbb{Z}_{2}\times\mathbb{Z}_{2}$ with non-trivial symmetry fractionalization in $\mathrm{H}^2(BG,\mathcal{A})$.
Here, for the toric code system, $\mathcal{A}$ is $\mathbb{Z}_{2}\times\mathbb{Z}_{2}$ generated by $e$- and $m$-anyons.
By interpolating each $\mathbb{Z}_{2}$-symmetry, we can construct a $T^{2}$-family of topologically ordered states,
and it is expected that this family defines a non-trivial family in $\mathrm{H}^2(T^{2},\mathcal{A})$.%
\footnote{
Generally, assuming that the $U_j$ symmetries individually act trivially on the category of anyons, we have the symmetry fractionalization class
\begin{equation}
    \eta \in {\rm H}^2(BG,\mathcal{A}).
\end{equation}
By associating each cycle $\theta_k$ of the $n$-torus to the corresponding $U_k$, we get a map
\begin{equation}
    U\colon T^n \to BG.
\end{equation}
Pulling back the class $\eta$, we obtain an invariant of the family
\begin{equation}
    U^*\eta \in {\rm H}^2(T^n,\mathcal{A}).
\end{equation}
The meaning of this invariant is that if we consider any 2-cycle $C \subset T^n$,
\begin{equation}
    \int_C U^*\eta \in \mathcal{A}
\end{equation}
labels the superselection sector of the particle-like defect obtained by wrapping parameters around $C$.
}

To demonstrate this, we consider the toric code model with $\mathbb{Z}_{2}\times\mathbb{Z}_{2}$ symmetry.
We put qubits on both links and plaquettes of a square lattice.
We study a modified Hamiltonian
\begin{equation}\label{eq:TC_m-anyon}
    H_{\text{I}}^{e}=-\sum_{v}A_{v}-\sum_{p}B_{p}-\sum_{p}\frac12 (1+X_{p}) \frac12 (1+B_{p}).
\end{equation}
where $A_{v}$ and $B_{p}$ are the usual vertex and plaquette terms of the toric code. This model is in the toric code phase, as one can see all its terms are minimized by taking a toric code ground state on edge spins along with $X_p = +1$ on plaquette spins.

This model has $\mathbb{Z}_{2}\times \mathbb{Z}_{2}$ symmetry generated by
\begin{equation}
    U_{1}^{e} = \prod_p X_{p},\;\; 
    U_{2}^{e} = \prod_{p} \prod_{i=1}^{4} \mathrm{CZ}_{p l_i},\label{eq: newU2_e_curcuit}
\end{equation}
where $\{l_{i}\}_{i=1}^{4}$ are adjacent links to the plaquette $p$ and $p$ is the control qubit for each CZ gate.
This symmetry exhibits a nontrivial symmetry fractionalization with respect to the $e$ anyon.
To see this, consider the symmetry acting partially on a rectangular region $M_{\square}$ with the smooth boundary:
\begin{equation}
    U_{1}^{e}(M_{\square}) = \prod_{p\in M_{\square}} X_{p},\;\;
    U_{2}^{e}(M_{\square}) = \prod_{p\in M_{\square}} \mathrm{CZ}_{p l_i}.
\end{equation}
Then, the commutation relation of them is given by
\begin{equation}\label{eq:frac_symmetry_e_commutation}
    U_{1}^{e}(M_{\square})U_{2}^{e}(M_{\square}) = (\prod_{l\in \partial M_{\square}} Z_{l}) U_{2}^{e}(M_{\square})U_{1}^{e}(M_{\square}),
\end{equation}
where $\partial M_{\square}$ is the set of links on the boundary of $M_{\square}$.
Therefore, if we define the representation $U_{M_{\square}}^{e}$ of $\mathbb{Z}_{2}\times \mathbb{Z}_{2}$ on $M_{\square}$ by
\begin{equation}\label{eq:frac_symmetry_e}
    U_{M_{\square}}^{e}(1,0) = 1,\;U_{M_{\square}}^{e}(1,0) = U_{1}^{e}(M_{\square}),\;U_{M_{\square}}^{e}(0,1) = U_{2}^{e}(M_{\square}),\;U_{M_{\square}}^{e}(1,1) = U_{2}^{e}(M_{\square})U_{1}^{e}(M_{\square}),
\end{equation}
then $\eta$ in \eqref{eq:sym_frac_gs} can be read off as 
\begin{equation}\label{eqetafrac}
    \eta_{\partial M}((g_1,g_2),(h_1,h_2))=
    \begin{cases*}
        \prod_{l\in \partial M_{\square}} Z_{l},\quad  &\text{if} $g_1=h_2=1$,\\
        1,\quad &\text{otherwise}.
    \end{cases*}
\end{equation}
This demonstrates the symmetry fractionalization principle described above: operators in the non-trivial superselection sector on the boundary may have a non-trivial action of the operator $\prod_{l \in \partial M_\square} Z_l$, which corresponds to transporting an $e$ anyon around the boundary.

Using the fact that abelian anyons $\{1,e,m,f\}$ in the toric code phase form a group $\mathbb{Z}_2\times\mathbb{Z}_2$, one obtains a nontrivial cocycle in $\mathrm{H}^2(\mathbb{Z}_2\times\mathbb{Z}_2,\mathbb{Z}_2\times\mathbb{Z}_2)$. 
Therefore, the $\mathbb{Z}_2\times\mathbb{Z}_2$ symmetry generated by $U_{1}^{e}$ and $U_{2}^{e}$ realizes a nontrivial symmetry fractionalization class that traps the $e$-anyon. 
We note that the cohomology class does not change if one redefine $U_{M_{\square}}^{e}(1,1)$ as $U_{1}^{m}(M)U_{2}^{m}(M)$. 

\begin{figure}[t]
    \centering
    \begin{minipage}{0.32\linewidth}
        \adjincludegraphics[page=1,scale=1.2,valign = c]{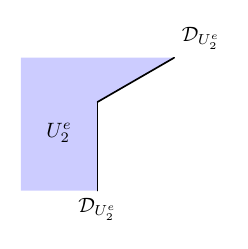}
      \caption*{(a)}
    \end{minipage}
    %\hfill
    %\hspace{-3pt}
    \begin{minipage}{0.32\linewidth}
        \adjincludegraphics[page=2,scale=1.2,valign = c]{figs/sym_frac_defect_1.pdf}
      \caption*{(b)}
    \end{minipage}
    %\hfill
    \begin{minipage}{0.32\linewidth}
        \adjincludegraphics[page=3,scale=1.2,valign = c]{figs/sym_frac_defect_1.pdf}
      \caption*{(c)}
    \end{minipage}
  
    \caption{
        Pictorial expression of the trapped anyon at the junction of the symmetry defect \eqref{eq:frac_symmetry_e}. 
        (a) Symmetry defect along the black line. This is realized by activating $U_{2}^{e}$ in the shaded area.
        (b) Upon acting $U_{1}^{e}$ in the shaded area, we obtain the configuration of the defects.
        (c) We rewrite the defect configuration of (b) by using \eqref{eq:frac_symmetry_e_commutation} and the definition \eqref{eq:frac_symmetry_e}. Now we have an $e$-anyon symmetry defect $\mathcal{D}_{e}$ at the junction of the defects.
    }
    \label{fig:sym_frac_defect}
\end{figure}

\paragraph{Trapped anyon at the junction of the symmetry defect}
To give another point of view why the $\mathbb{Z}_2\times\mathbb{Z}_2$ symmetry with the nontrivial symmetry fractionalization class is related to $e$-anyon, let us consider the symmetry defects of the $\mathbb{Z}_2\times\mathbb{Z}_2$ symmetry. Recall that the symmetry defect along a line is realized by acting the corresponding symmetry operator on one of the half areas separated by the line, see Fig.~\ref{fig:sym_frac_defect} (a).
We denote the symmetry defect for an operator $U$ by $\mathcal{D}_{U}$.
Let us consider the configuration of three defects $U_{1}^{e},U_{2}^{e},U_{2}^{e}U_{1}^{e}$ as in Fig.~\ref{fig:sym_frac_defect} (c), which is realized by two sequential actions of $U_{2}^{e}$ and $U_{1}^{e}$ as Fig.~\ref{fig:sym_frac_defect} (a)(b). Though the symmetry defect $\mathcal{D}_{U_{1}^{e}}\mathcal{D}_{U_{2}^{e}}$ is almost likely to the defect $\mathcal{D}_{U_{2}^{2}U_{1}^{e}}$, they are not completely the same. Due to the commutation \eqref{eq:frac_symmetry_e_commutation} and the definition \eqref{eq:frac_symmetry_e}, the former is related to the latter up to the action of the $e$-anyon string end at the junction, which corresponds to the symmetry defect for the one-form symmetry generated by the $e$-anyon line. Therefore, the $e$-anyon is trapped (attached) to the junction of the defects Fig.~\ref{fig:sym_frac_defect} (c).

\subsection{\texorpdfstring{$2$}{2}-parameter family}

Let us interpolate the symmetry to construct a $T^{2}$-family of topologically ordered states. We define the interpolation of $U_{1}^{e}$ and $U_{2}^{e}$ as
\begin{align}\label{eq: U_e interpolation}
\begin{aligned}
    U_{1}^{e}(\theta) &\coloneqq \prod_{p} e^{i\frac{\theta}{4}(1-X_{p})},\\
    U_{2}^{e}(\theta) &\coloneqq 
    \prod_{l} e^{i\frac{\theta}{4}(1-\mathrm{CZ}_{p_{1}l}\mathrm{CZ}_{lp_{2}})},
\end{aligned}
\end{align}
where $p_{1},p_{2}$ are the two plaquettes adjacent to the link $l$.
Then, we can define a $T^{2}$-family of topologically ordered systems by
\begin{equation}\label{eqnT2family}
    H_{\text{I}}^{e}(\theta_{1},\theta_{2})\coloneqq U_{1}^{e}(\theta_{1})U_{2}^{e}(\theta_{2})H_{\text{I}}^{e}U_{2}^{e}(\theta_{2})^{\dagger}U_{1}^{e}(\theta_{1})^{\dagger}.
\end{equation}
To have a $T^2$-family, we want
\begin{equation}
    \begin{split}
        H_{\rm I}^e(\theta_1,2\pi) = H_{\rm I}^e(\theta_1,0),\\
        H_{\rm I}^e(2\pi,\theta_2) = H_{\rm I}^e(0,\theta_2).
    \end{split}
\end{equation}
Since we have placed $U_2^e$ on the inside of \eqref{eqnT2family}, the first is guaranteed by $U_2^e(2\pi)$ being a symmetry. For the second, we need $U_1^e(2\pi)$ to be a symmetry and for it also to commute with $U_2^e(\theta)$ (see the discussion around \eqref{eqnTnfamily}). One checks that this holds with this choice of interpolation.

It may be helpful to compare this trapped anyon picture with the $em$-exchange symmetry and the corresponding $S^1$-family in Section~\ref{sec: em pump}. The main idea of constructing the $S^1$-parameter family in Section~\ref{sec: 1-para} is to interpolate the $em$-exchange symmetry, which gives the $em$-exchange defect when considering the partial symmetry action, hence we also get a pump of the $em$-exchange defect, a nontrivial codimension-1 defect, when we adiabatically traverse the $S^1$. In contrast, we saw above that the intersection of $\mathbb{Z}_2^2$ defects traps an $e$ anyon, a nontrivial codimension-2 defect. Thus, we expect that a certain parameter texture for this family (like a Skyrmion) will trap the $e$ anyon as well. We are not able to show rigorously, so to analyze the non-triviality of the family, we will give an alternative characterization in terms of the boundary algebras.

\subsection{Non-triviality of the family}

In this section, we demonstrate that the constructed $T^{2}$-family is nontrivial. 
To this end, we consider a system with a boundary and study the pumping phenomenon that occurs at the boundary.

\begin{figure}[t]
    \centering
        \adjincludegraphics[scale=1.25,trim={10pt 10pt 10pt 10pt},valign = c]{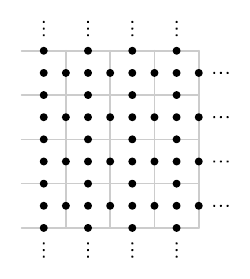}
    \caption{
    Underlying lattice of the truncated Hamiltonian. The left side represents the boundary, and each black dot denotes a local Hilbert space.
    }
    \label{fig: T2truncation}
\end{figure}

\paragraph{Boundary algebra for the rough boundary} Now we wish to truncate the Hamiltonian such that we obtain a $T^2$-family of boundary algebras. We will choose a rough boundary with plaquette spins occupying the boundary plaquettes between the perpendicular boundary edges (see Fig.~\ref{fig: T2truncation}). 
The symmetries are chosen so that the product over edges and plaquettes in \eqref{eq: U_e interpolation} is taken to include all of the edges and plaquettes in the boundary geometry. Call these $U_1^{e,r}(\theta_1)$ and $U_2^{e,r}(\theta_2)$. One checks that with this choice
\begin{equation}
    U_1^e(2\pi) = \prod_p X_p
\end{equation}
commutes with $U_2^e(\theta)$, and
\begin{equation}
    U_2^e(2\pi) = \prod_l CZ_{p_1 l} CZ_{p_2 l}.
\end{equation}

If we consider the Hamiltonian as in \eqref{eq:TC_m-anyon}, taking the $A_v$ and $B_p$ terms which fit in the lattice as usual, and also the new terms which fit in the lattice, then we can check that the unitaries above are symmetries of the truncated Hamiltonian $H^r$. Then
\begin{equation}\label{eqnT2familyunitary}
    H_{\rm I}^{e,r}(\theta_{1},\theta_{2})\coloneqq U_{1}^{e,r}(\theta_{1})U_{2}^{e,r}(\theta_{2})H_{\text{I}}^{e,r}U_{2}^{e,r}(\theta_{2})^{\dagger}U_{1}^{e,r}(\theta_{1})^{\dagger}.
\end{equation}
defines a $T^2$-family of commuting projectors with boundary.

The boundary algebra for $H_{\rm I}^{e}$ may be computed by noting that the plaquette spin projectors are redundant with the toric code plaquette projector $\frac12(1+B_p)$, so as a net of projections we can take the local plaquette spin generator to be $\frac12(1+X_p)$ which just projects it out. We have not included these projections for the boundary plaquettes, so the boundary algebra $\mathcal{A}^r(0,0)$ for $\theta_1=\theta_2=0$ is generated by:
\begin{equation}
    \Bigg\{\;\;
    \adjincludegraphics[scale=1.25,trim={10pt 10pt 10pt 10pt},valign = c]{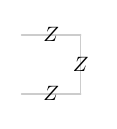},\;\;
    \adjincludegraphics[scale=1.25,trim={10pt 10pt 10pt 10pt},valign = c]{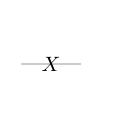},\;\;
    \adjincludegraphics[scale=1.25,trim={10pt 10pt 10pt 10pt},valign = c]{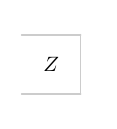},\;\;
    \adjincludegraphics[scale=1.25,trim={10pt 10pt 10pt 10pt},valign = c]{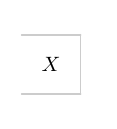}\;\;
    \Bigg\}
\end{equation}
We call these operators $\tilde Z_{l_1} \tilde Z_{l_2}$, $\tilde X_l$, $\tilde Z_p$, and $\tilde X_p$, respectively, embedding them as the $\prod_l \tilde X_l$-symmetric subalgebra of a spin-$1/2$ chain with two sublattices consisting of $l$ and $p$ spins. The other boundary algebras $\mathcal{A}^r(\theta_1,\theta_2)$ may be obtained by applying the unitaries defining the family in \eqref{eqnT2familyunitary}:
\begin{equation}
    \mathcal{A}^r(\theta_1,\theta_2) = U_{1}^{e,r}(\theta_{1})U_{2}^{e,r}(\theta_{2})\mathcal{A}^r(0,0)U_{2}^{e,r}(\theta_{2})^{\dagger}U_{1}^{e,r}(\theta_{1})^{\dagger}.   
\end{equation}
The notation means that the elements of $\mathcal{A}^r(\theta_1,\theta_2)$ are operators of the form
\begin{align}
\begin{split}
    &U_{1}^{e,r}(\theta_{1})U_{2}^{e,r}(\theta_{2}) P \,a\, P U_{2}^{e,r}(\theta_{2})^{\dagger}U_{1}^{e,r}(\theta_{1})^{\dagger} \\
    &= P(\theta_1,\theta_2)U_{1}^{e,r}(\theta_{1})U_{2}^{e,r}(\theta_{2}) \,a\, U_{2}^{e,r}(\theta_{2})^{\dagger}U_{1}^{e,r}(\theta_{1})^{\dagger} P(\theta_1,\theta_2),
\end{split}
\end{align}
where $P(\theta_1,\theta_2) = U_{1}^{e,r}(\theta_{1})U_{2}^{e,r}(\theta_{2}) P  U_{2}^{e,r}(\theta_{2})^{\dagger}U_{1}^{e,r}(\theta_{1})^{\dagger}$ is the ground state projector for $H_{\rm I}^{e,r}(\theta_1,\theta_2)$ and $P$ is the ground state projector for $H_{\rm I}^{e,r}(0,0)$.

Periodicity of the family implies $\mathcal{A}^r(\theta_1,2\pi) = \mathcal{A}^r(\theta_1,0)$ and $\mathcal{A}^r(2\pi,\theta_2)=\mathcal{A}^r(0,\theta_2)$, but the automorphism induced by $U_2^{e,r}$ and $U_1^{e,r}$ may be non-trivial. A short calculation yields
\begin{align}
    \tilde U_1^{e,r}: \begin{cases}
    \tilde Z_{l_1} \tilde Z_{l_2} \mapsto \tilde Z_{l_1} \tilde Z_{l_2} \\
    \tilde X_l \mapsto \tilde X_l \\
    \tilde X_p \mapsto \tilde X_p \\
    \tilde Z_p \mapsto - \tilde Z_p
\end{cases} \\
    \tilde U_2^{e,r}: \begin{cases}
    \tilde Z_{l_1} \tilde Z_{l_2} \mapsto \tilde Z_{l_1} \tilde Z_{l_2} \\
    \tilde X_l \mapsto \tilde Z_{p_1} \tilde X_l \tilde Z_{p_2} \\
    \tilde X_p \mapsto \tilde Z_{l_1} \tilde X_p \tilde Z_{l_2} \\
    \tilde Z_p \mapsto \tilde Z_p
\end{cases}
\end{align}
where $p_1,p_2$ refers to the plaquettes to the left and right of $l$, and $l_1, l_2$ refers to the edges to the left and right of $p$.

Both of these automorphisms are locally generated in the boundary algebra, namely
\begin{align}
    \begin{split}
        \tilde U_1^{e,r} &= \prod_p \tilde X_p, \\
        \tilde U_2^{e,r} &= \prod_p C\tilde Z_{l_1 p} C\tilde Z_{p l_1},
    \end{split}
\end{align}
where we grouped the second product so that $C\tilde Z_{l_1 p} C\tilde Z_{p l_1}$ is in the boundary algebra. Note $C\tilde Z_{l_1 p}$ alone is \emph{not} in the boundary algebra, since it does not commute with $\prod \tilde X_l$. Thus, unlike the $em$ swap symmetry, they are trivial QCAs. This is partially to be expected, since the bulk symmetries do not permute the anyons, but in principle there could be a translation part as well, although we do not observe that here.

What makes the family non-contractible from this point of view? Although the automorphisms $\tilde U_1^{e,r}$ and $\tilde U_2^{e,r}$ are locally-generated, consider applying $\tilde U_2^{e,r}$ to a region, where we number boundary plaquettes with even integers $p$ and boundary edges with odd integers $l$:
\begin{equation}
    \tilde U_2^{e,r,[0,2n]}= \prod_{0 \le p \le 2n} C\tilde Z_{l_1 p} C\tilde Z_{p l_1}.
\end{equation}
This operator does not commute with $\tilde U_1^{e,r}$, instead we find
\begin{equation}
    \tilde U_1^{e,r}\tilde U_2^{e,r,[0,2n]} (\tilde U_1^{e,r})^\dagger (\tilde U_2^{e,r,[0,2n]})^\dagger = \tilde Z_{-1} \tilde Z_{2n+1}.
\end{equation}
These endpoint operators do not factorize in the boundary algebra, $\tilde Z_l$ is in the non-trivial $e$ superselection sector, and we claim without proof that this captures the $e$ anyon pumping we argued intuitively in the bulk.\footnote{Indeed, this is nothing but the 2-cocycle $\eta_{\partial M}$ evaluated on the torus.}

Indeed, the superselection sector of this endpoint operator in the commutator is quantized, and therefore invariant under deformations of the family, since these would lead to deformations of $\tilde U_1^{e,r}$ and $\tilde U_2^{e,r}$ such that globally they always commute. Thus we find a connection between the non-contractibility of the bulk $T^2$-family and a boundary charge pump.

This calculation has a natural interpretation in terms of the homotopy theory of QCA. Indeed, we expect that as in \cite{czajka2025anomalieslatticehomotopyquantum} there will be a classifying space $\mathcal{Q}(\mathcal{A}^r)$ of QCA in the boundary algebra $\mathcal{A}^r$, such that our $T^2$-family has a classifying map $T^2 \to \mathcal{Q}(\mathcal{A}^r)$. Above we found that the map on $\pi_1 T^2$ is trivial, since the corresponding QCA are locally generated. Therefore, the only remaining invariant of the $T^2$-family is the degree, an element of $\pi_2 \mathcal{Q}(\mathcal{A}^r)$. Unlike a tensor product algebra, this homotopy group is non-trivial, since there are local projective unitaries which are not generated by elements of the boundary algebra, such as ${\rm Ad}_{\tilde Z_{-1}}$ above. These correspond to the superselection sectors of the algebra, hence our $T^2$-family has an invariant in $\pi_2 \mathcal{Q}(\mathcal{A}^r) = \mathbb{Z}_2 \times \mathbb{Z}_2$. The commutator above can be identified with this invariant by the methods in \cite{czajka2025anomalieslatticehomotopyquantum}. We leave a formal proof of this to future work.

Let us give a complementary point of view on this boundary pump which allows us to think of it as a ``pump of pumps''. By comparing the boundary algebras at $\theta_{2}=0$ and $\theta_{2}=2\pi$, we obtain, for each $\theta_{1}$, an $\mathbb{Z}_{2}$-symmetric automorphism of the boundary algebra (the $\mathbb{Z}_2$ symmetry being given by $\tilde U_1^{e,r}$). 
This family of automorphisms acting on the space of boundary conditions is given by
\begin{equation}
    \widetilde{Z}_{l_{1}}\widetilde{Z}_{l_{2}}\mapsto \widetilde{Z}_{l_{1}}\widetilde{Z}_{l_{2}},\;\;
    \widetilde{X}_{l}\mapsto\widetilde{Z}_{p_{1}}^{\theta_{1}}\widetilde{X}_{l}\widetilde{Z}_{p_{2}}^{\theta_{1}},\;\;
    \widetilde{X}_{p}\mapsto\widetilde{Z}_{l_{1}}\widetilde{X}_{p}\widetilde{Z}_{l_{2}},\;\;
    \widetilde{Z}_{p}^{\theta_{1}}\mapsto\widetilde{Z}_{p}^{\theta_{1}}.
\end{equation}
where $l_{1}$ and $l_{2}$ are the two links adjacent to the plaquette $p$ on the boundary and $p_{1}$ and $p_{2}$ are the two plaquettes adjacent to the link $l$ on the boundary, 
and $\widetilde{Z}^{\theta}\coloneq e^{i\frac{\theta}{4}\widetilde{X}}\widetilde{Z}e^{-i\frac{\theta}{4}\widetilde{X}}$. This automorphism is implemented by the following MPO tensors:
\begin{equation}
    B(\theta_{1}) = \adjincludegraphics[scale=1.25,trim={10pt 10pt 10pt 10pt},valign = c]{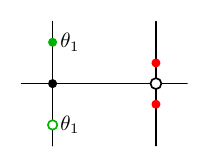},
\end{equation}
where left vertical line is the physical leg actiog on a plaquette effective spin and right vertical line is the physical leg acting on a link effective spin, and horizontal lines are virtual legs of MPO.
The black and white dot is the copy and multiplication tensors in $\mathbb{C}[\mathbb{Z}_{2}]$.
The red dots immediately above and below each vertex denote Hadamard gates.
The filled green tensor represents $e^{i\frac{\theta_{1}}{4}\widetilde{X}_{p}}$, and the green unfilled dot represents $e^{-i\frac{\theta_{1}}{4}\widetilde{X}_{p}}$.
We can regard this MPO as an $S^{1}$-family of $\mathbb{Z}_{2}$-symmetric matrix product states by folding the MPO.
Then, we can show that this $S^{1}$-family has a nontrivial topological index by computing the $\mathbb{Z}_{2}$-invariant defined in \cite{Shiozaki:2021weu}.
Thus, this is a nontrivial family of automorphisms of the boundary algebra, and we conclude that the $T^{2}$-family $H_{\rm I}^{e}(\theta_{1},\theta_{2})$ is nontrivial.

\paragraph{Pump of $S^{1}$-paramterized family of SPT states}

In the above discussion, we analyzed the pumping phenomenon by focusing on automorphisms of the boundary algebra.
However, because the physical meaning of the boundary algebra itself is not entirely clear, it is not evident what charge is pumped as a consequence of the nontriviality of the family.
To provide a more heuristic but physical argument, we present a description of the pump of invertible states.

Recall that the action of the boundary algebra for $H_{\rm I}^{e,r}(\theta_{1},\theta_{2}=0)$ on the space of boundary conditions is given by
\begin{equation}
    \widetilde{Z}_{l_{1}}\widetilde{Z}_{l_{2}},\;\;\widetilde{X}_{l},\;\;\widetilde{X}_{p},\;\;\widetilde{Z}_{p}^{\theta_{1}},
\end{equation}
with $\mathbb{Z}_{2}$-symmetry generated by $\prod_{l}\widetilde{X}_{l}$.
Using these, we construct a family of $\mathbb{Z}_{2}$-symmetric Hamiltonian that is $2\pi$-periodic in $\theta_{1}$ and whose ground state is one-dimensional, that is , an $S^{1}$-family of $\mathbb{Z}_{2}$ SPT states.
The simplest choice is
\begin{equation}
    \widetilde{H}_{I,0}^{e,r}(\theta_{1}) \coloneqq -\sum_{l}\widetilde{X}_{l}-\sum_{p}\widetilde{X}_{p}.
\end{equation}
As an $S^{1}$-family of $\mathbb{Z}_{2}$-symmetric invertible states, this is trivial. 
However, after varying $\theta_{2}$ from $0$ to $2\pi$, the Hamiltonian is transformed as
\begin{equation}
    \widetilde{H}_{I,1}^{e,r}(\theta_{1}) = -\sum_{l}\widetilde{Z}_{p_{1}}^{\theta_{1}}\widetilde{X}_{l}\widetilde{Z}_{p_{2}}^{\theta_{1}}-\sum_{p}\widetilde{Z}_{l_{1}}\widetilde{X}_{p}\widetilde{Z}_{l_{2}}.
\end{equation}
As an $S^{1}$-family of $\mathbb{Z}_{2}$-symmetric invertible states, this is non-trivial \cite{Shiozaki:2021weu}.
Since $\widetilde{H}_{I,0}^{e,r}(\theta_{1})$ and $\widetilde{H}_{I,1}^{e,r}(\theta_{1})$ have different indices as $\mathbb{Z}_{2}$-symmetric families, they cannot be connected by local deformations.
This means that the nontrivial $S^{1}$family of $\mathbb{Z}_{2}$-symmetric invertible states is pumped to the boundary by varying $\theta_{2}$,
and implies that the $T^{2}$-family $H_{\rm I}^{e}(\theta_{1},\theta_{2})$ is nontrivial.

\subsection{Other \texorpdfstring{$T^{2}$}{T2}-families}

Similar to the above construction, we can also construct a $T^{2}$-family that traps the $m$-anyon.
To this end, we put qubits on both vertices and links of a square lattice.
Then, the Hamiltonian is defined by
\begin{equation}
    H_{\text{I}}^{m}=-\sum_{v}A_{v}-\sum_{p}B_{p}-\sum_{v}\frac{1}{2}(1+Z_{v})\frac{1}{2}(1+A_{v}).
\end{equation}
This model has $\mathbb{Z}_{2}\times \mathbb{Z}_{2}$ symmetry generated by
\begin{equation}
    U_{1}^{m} = \prod_v Z_{v},\;\; U_{2}^{m} = \prod_{v} \prod_{i=1}^{4} H_{l_i}\mathrm{CNOT}_{v l_i}H_{l_i},
\end{equation}
where $\{l_{i}\}_{i=1}^{4}$ are adjacent links to the vertex $v$ and $v$ is the target qubit for each CNOT gate, that is, $\mathrm{CNOT}_{v l_i}=e^{i\frac{\pi}{4}(I-Z_{l_{i}})(I-X_{v})}$.

We can take the interpolation of $U_{1}^{m}$ and $U_{2}^{m}$ as 
\begin{equation}
    U_{1}^{m}(\theta) \coloneq \prod_v e^{i\frac{\theta}{4}(1-Z_{v})},\;\; 
    U_{2}^{m}(\theta) = \prod_{v} \prod_{i=1}^{4} e^{i\frac{\theta}{4}(1-H_{l}\mathrm{CNOT}_{v_{1} l}\mathrm{CNOT}_{v_{2} l}H_{l})},
\end{equation}
where $v_{1}$ and $v_{2}$ are the two vertices adjacent to the link $l$.
Then, we can define a $T^{2}$-family of topologically ordered systems by
\begin{equation}
    H_{\rm I}^{m}(\theta_{1},\theta_{2})\coloneqq U_{1}^{m}(\theta_{1})U_{2}^{m}(\theta_{2})H_{\text{I}}^{m}U_{2}^{m}(\theta_{2})^{\dagger}U_{1}^{m}(\theta_{1})^{\dagger}.
\end{equation}

The boundary algebra for the smooth boundary is generated by the following stabilizers:
\begin{equation}
    \Bigg\{\;\;
    \adjincludegraphics[scale=1.25,trim={10pt 10pt 10pt 10pt},valign = c]{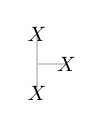},\;\;
    \adjincludegraphics[scale=1.25,trim={10pt 10pt 10pt 10pt},valign = c]{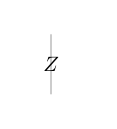},\;\;
    \adjincludegraphics[scale=1.25,trim={10pt 10pt 10pt 10pt},valign = c]{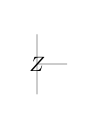},\;\;
    \adjincludegraphics[scale=1.25,trim={10pt 10pt 10pt 10pt},valign = c]{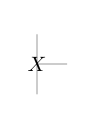}\;\;
    \Bigg\},
\end{equation}
with global $\mathbb{Z}_{2}$-symmetry generated by $\prod X_{l}$ on the boundary. 
We can also show that the $T^{2}$-family constructed by interpolating $U_{1}^{m}$ and $U_{2}^{m}$ is nontrivial by the same argument as above.

By combining $e$- and $m$-trapped families, we can also construct a $T^{2}$-family that traps the $f$-anyon.
We put qubits on vertices, links, and plaquettes of a square lattice.
Then, the Hamiltonian is defined by
\begin{equation}
    H_{\text{I}}^{f}=-\sum_{v}A_{v}-\sum_{p}B_{p}-\sum_{v}\frac{1}{2}(1+Z_{v})\frac{1}{2}(1+A_{v})-\sum_{p}\frac{1}{2}(1+X_{p})(1+B_{p}).     
\end{equation}
This model has $\mathbb{Z}_{2}\times \mathbb{Z}_{2}$ symmetry generated by
\begin{equation}
    U_{1}^{f} = \prod_v Z_{v}\prod_p X_{p},\;\; U_{2}^{f} = \prod_{v} \prod_{i=1}^{4} H_{l_i}\mathrm{CNOT}_{v l_i}H_{l_i}\prod_{p} \prod_{i=1}^{4} \mathrm{CZ}_{p l_i}.
\end{equation}
This symmetry has a nontrivial symmetry fractionalization class that traps the $f$-anyon.

We can take the interpolation of $U_{1}^{f}$ and $U_{2}^{f}$ as
\begin{equation}
    U_{1}^{f}(\theta) \coloneq U_{1}^{e}(\theta)U_{1}^{m}(\theta),\;\; 
    U_{2}^{f}(\theta) \coloneq U_{2}^{e}(\theta)U_{2}^{m}(\theta),
\end{equation}
and define a $T^{2}$-family of topologically ordered systems by
\begin{equation}
    H_{\rm I}^{f}(\theta_{1},\theta_{2})\coloneqq U_{1}^{f}(\theta_{1})U_{2}^{f}(\theta_{2})H_{\text{I}}^{f}U_{2}^{f}(\theta_{2})^{\dagger}U_{1}^{f}(\theta_{1})^{\dagger}.
\end{equation}
One can show that the $T^{2}$-family constructed by interpolating $U_{1}^{f}$ and $U_{2}^{f}$ is nontrivial by the same argument as above. 

\subsection{Other interpolations: Klein bottle family}
By using other interpolation of $\mathbb{Z}_{2}\times\mathbb{Z}_{2}$-symmetry, we can also construct a parameterized family of the toric code phase over the Klein bottle.
Let us consider the Hamiltonian with plaquette qubits:
\begin{equation}\label{eq:TC_e-anyon}
    H_{\text{II}}^{e}=-\sum_{v}A_{v}-\sum_{p}B_{p}-\sum_{p}\frac{1}{2}(1+X_{p})\frac{1}{2}(1+B_{p}).
\end{equation}
This is the same as the Hamiltonian \eqref{eq:TC_m-anyon} and consider the $\mathbb{Z}_2\times\mathbb{Z}_2$ symmetry defined in \eqref{eq: newU2_e_curcuit}
\begin{equation}
    U_{1}^{e} = \prod_p X_{p},\;\; 
    U_{2}^{e} = \prod_{p} \prod_{i=1}^{4} \mathrm{CZ}_{p l_i}.
\end{equation}
So far, this is the same as the previous model $H_{\rm I}^{e}$, but we consider a different interpolation of $U_{2}^{e}$:
\begin{align}\label{eq: U_e prime interpolation}
    U_{1}^{\prime e}(\theta) &\coloneq \prod_{p} e^{i\frac{\theta}{4}(1-X_{p})},\\
    U_{2}^{\prime e}(\theta) &\coloneqq \prod_{l} e^{i\frac{\theta}{8}(I-Z_{p_{1}})(I-Z_{l})-i\frac{\theta}{8}(I-Z_{l})(I-Z_{p_{2}})},\nonumber\\
    &=\prod_{l} e^{-i\frac{\theta}{8}(Z_{p_{1}}-Z_{p_{2}})(I-Z_{l})}.
\end{align}
By using these interpolations, we can define a $2$-parameter family of topologically ordered systems by
\begin{gather}
    H_{\text{II}}^{e}(\theta_1,\theta_2)\coloneq U_{1}^{e\prime}(\theta_1)U_{2}^{e\prime}(\theta_2)~H_{\text{II}}^{e}~U_{2}^{e\prime}(\theta_2)^\dagger U_{1}^{e\prime}(\theta_2)^\dagger, \label{eq:TC_family_e}.
\end{gather}

The essential difference from $H_{\rm I}^{e}$ is the parametrization of $H_{\text{II}}^{e}(\theta_1,\theta_2)$.
In fact, $H_{\text{II}}^{e}(\theta_1,\theta_2)$ satisfies
\begin{equation}
    H_{\text{II}}^{e}(\theta_1+2\pi,\theta_2)=H_{\text{II}}^{e}(\theta_1,-\theta_2),\quad H_{\text{II}}^{e}(\theta_1,\theta_2+2\pi)=H_{\text{II}}^{e}(\theta_1,\theta_2),
\end{equation}
due to the mild non-commutativity of $U_{1}^{e\prime}$ and $U_{2}^{e\prime}(\theta)$.
Therefore, $H_{\text{II}}^{e}(\theta_1,\theta_2)$ is a family over the Klein bottle $K$.
However, we can follow the same argument as above and show that this family is also nontrivial by considering the boundary algebra.

\section{Crystalline Toric Code Model: higher order anyon pump}
\label{sec: crystalline pump}

In this section, we discuss an $S^1$-family in the toric code phase which is non-contractible in the presence of a rotation symmetry. It has the feature of pumping an anyon to each corner of the system over a $2\pi$ period of the parameter.

\subsection{Hamiltonian}
\label{sec: crystalline Ham}
On each edge and plaquette, we put $\mathbb{C}^{2}$ as a local Hilbert space, 
and let $\sigma^{x}$ and $\sigma^{z}$ be the Pauli matrices acting on the plaquette spins,
and let $X$ and $Z$ be the Pauli matrices acting on the edge spins. 
Then, the stabilizers are defined as follows:
\begin{equation}\label{eq: crystalline stabilizers}
    A_{v}\coloneq\adjincludegraphics[scale=1.2,trim={10pt 10pt 10pt 10pt},valign = c]{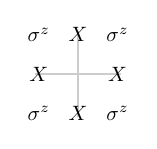},\;\;
    B^{(1)}_{p}\coloneq\adjincludegraphics[scale=1.2,trim={10pt 10pt 10pt 10pt},valign = c]{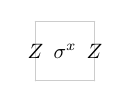},\;\;
    B^{(2)}_{p}\coloneq\adjincludegraphics[scale=1.2,trim={10pt 10pt 10pt 10pt},valign = c]{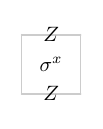}.
\end{equation}
Note that combining $B^{(1)}_{p}$ and $B^{(2)}_{p}$ gives the usual toric code plaquette operator $B_{p}$.
Let $H_{I}$ be the corresponding Hamiltonian:
\begin{equation}\label{eq: e anyon pump Hamiltonain}
    H_{I}= -\sum_{v} A^{e}_{v} -\sum_{p} B^{e,(1)}_{p} -\sum_{p} B^{e,(2)}_{p}.
\end{equation}
This model has $\mathbb{Z}_{2}$ onsite symmetry generated by
\begin{equation}
    U_{e}\coloneq\prod_{p} \sigma^{x},
\end{equation}
and crystalline $C_{4}$-symmetry around vertex and plaquette centers.

We note that this model has the same anyonic excitations as the toric code.
For a loop $\gamma_{e}$, the $e$-anyon loop operator is defined by
\begin{equation}
    W_e(\gamma_e) = \prod_{l \in \gamma_e} Z_l.
\end{equation}
This is the same as in the original toric code.
By contrast, the loop operator for the $m$-anyon requires a slight modification:
For a loop $\gamma_{m}$, the original $m$-anyon loop operator
\begin{equation}
    W_m(\gamma_m) = \prod_{l \cap \gamma_m \neq \varnothing} X_l
\end{equation}
does not commute with the Hamiltonian and is therefore not topological.
To obtain a topological operator, we put $\sigma^{z}$ at the corners of the loop:
\begin{equation}\label{eq: m-anyon crystalline}
    W'_m(\gamma_m) = \prod_{l \cap \gamma_m \neq \varnothing} X_l \prod_{p \in \mathrm{cor(\gamma_m)}}\sigma^{z}_{p},
\end{equation}
where $\mathrm{cor(\gamma_m)}$ denotes the set of plaquettes at the corners of the loop $\gamma_{m}$.
For example, the vertex term $A_{v}$ in \eqref{eq: crystalline stabilizers} is the minimal loop operator encircling the vertex $v$.%
\footnote{
This modification can be naturally understood from the gauging point of view.
We return to this point in Section~\ref{sec: nontriviality anyon pump} after introducing the gauged model.}

\subsection{Ground state}

The ground state of the model is realized as a simultaneous eigenstate of all terms with eigenvalue $1$.
To construct this state, we first consider an eigenstate of $B_{p}=B_{p}^{(1)}B_{p}^{(2)}$.
Such states are given by the following configuration:
\begin{equation}
    \adjincludegraphics[scale=1.75,trim={10pt 10pt 10pt 10pt},valign = c]{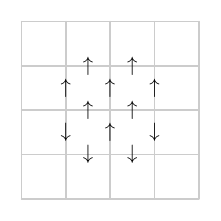}
\end{equation}
For a given configuration, $B_{p}^{e,(1)}$ (or equivalently $B_{p}^{e,(2)}$) determines the plaquette spin configurations uniquely:
\begin{equation}
    \adjincludegraphics[scale=1.75,trim={10pt 10pt 10pt 10pt},valign = c]{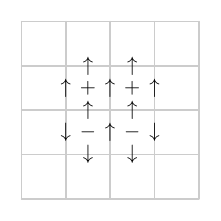}
\end{equation}
This state is not an eigenstate of $A_{v}^{e}$, but this term preserve the configuration that minimizes $\{B_{p}^{e,(i)}\}_{i=1,2}$ terms. 
Therefore, the ground state is an equal-weight superposition of the configuration:
\begin{equation}
    \ket{\mathrm{G.S.}}=\prod_{v}\frac{1+A_{v}^{e}}{2}\ket{0},
\end{equation}
where $\ket{0}$ is a reference state defined by $\ket{0}\coloneqq \bigotimes_{p}\ket{+}\bigotimes_{v}\ket{\uparrow}$.

The ground state degeneracy is the same as that of the toric code, and in particular, the degeneracy on the plane is one.  
In fact, although we have introduced an additional $\mathbb{C}^2$ degree of freedom on each plaquette in addition to those of the toric code, 
this extra degree of freedom is completely fixed by the operators $B_p^{e,(1)}$ and $B_p^{e,(2)}$, as we have already seen.  
Therefore, the dimension of the ground state degeneracy coincides with that of the toric code.

Although the superselection theory is the same, the minimum energy excitations are slightly different from those in the toric code.  
First, as in the toric code, there exist $e$-anyons and $m$-anyons.
Indeed, one can create anyon excitations by acting with $Z$ or $X$ operators on the ground state.
\footnote{
    However, the string of an $m$ anyon is slightly different from that in the usual toric code. 
    In the usual toric code, it suffices to apply $X$ operators along the plaquettes, whereas in this model it is also necessary to apply $\sigma^{x}$ on the plaquette at the corner where the string bends by $90^{\circ}$, if we do not want to create further excitations there.
}  
In addition to these, one can also create an anyon excitation by applying $\sigma^z$ to a plaquette. 
However, since the ground state stabilizes both $B_p^{e,(1)}$ and $B_p^{e,(2)}$, the action of $\sigma^z$ is nothing but the product of two adjacent $Z$ operators.  
Therefore, this excitation is equivalent to a pair of two $e$ anyons.
One can also create an excitation by applying the $\sigma^x$ operator to a plaquette.  
However, this excitation is not topological and does not change the ground state degeneracy.
Therefore, we find that this model has exactly the same anyon excitations as the toric code.

\subsection{Partial symmetry action}
\begin{figure}[t]
    \centering
    \begin{minipage}{0.32\linewidth}
        \adjincludegraphics[scale=1.5,trim={10pt 10pt 10pt 10pt},valign = c]{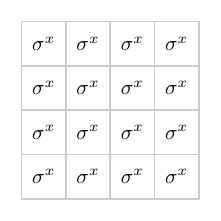}
      \caption*{(a)}
    \end{minipage}
    \hfill
    \begin{minipage}{0.32\linewidth}
        \adjincludegraphics[scale=1.5,trim={10pt 10pt 10pt 10pt},valign = c]{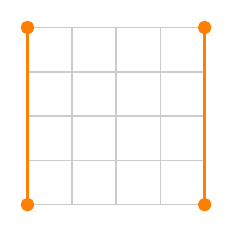}
      \caption*{(b)}
    \end{minipage}
    \hfill
    \begin{minipage}{0.32\linewidth}
        \adjincludegraphics[scale=1.5,trim={10pt 10pt 10pt 10pt},valign = c]{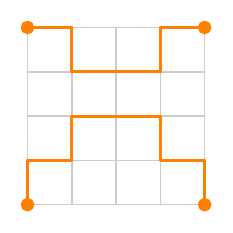}
      \caption*{(c)}
    \end{minipage}
  
    \caption{
        Partial symmetry action of $U_{e}$ to the rectangular region.
        The orange lines represent the $e$-anyon strings and the orange dots represent the $e$-anyons.
        (a) The action of $U_{e}$ to the rectangular region.
        (b) Replacing $\sigma^{x}$ with $Z$ operators using $B_p^{e,(1)}$ only.
        (c) Replacing $\sigma^{x}$ with $Z$ operators using $B_p^{e,(1)}$ and $B_p^{e,(2)}$.
        In both cases, $e$-anyons appear at the corners.
    }
    \label{fig:symmetry_action}
\end{figure}

Let us consider the partial symmetry action of $U$ to the rectangular region.
As already mentioned, the action of $\sigma^{x}$ on the ground state can be replaced by the action of $Z$ on the left/right of the plaquette, or on the top/bottom, depending on whether one uses $B_p^{e,(1)}$ or $B_p^{e,(2)}$.
By appropriately replacing each $\sigma^{x}$ with a $Z$ operator using either $B_p^{e,(1)}$ or $B_p^{e,(2)}$, an $e$-anyon excitation appears at the corners, regardless of the choice between $B_p^{e,(1)}$ and $B_p^{e,(2)}$ (see Fig.~\ref{fig:symmetry_action}).

\subsection{Disentangler}
If we break the $C_{4}$ symmetry, we can decouple the toric code and the plaquette degrees of freedom.
% \delete{In fact, by applying the following disentangler, we can decouple the system into the toric code and the plaquette spins:}
Indeed, we can construct the disentanger of the SET Hamiltonian \eqref{eq: e anyon pump Hamiltonain} as follows. First, we define the local circuit as
\begin{equation}\label{eq: disentanlger C_4}
    \adjincludegraphics[scale=1.2,trim={10pt 10pt 10pt 10pt},valign = c]{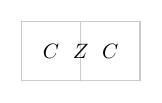}.
\end{equation}
Here, $C$ denotes the control bit in the $Z$-basis, and the disentangler means applying a $Z$ operator on the link between two plaquette operators when they are oppositely oriented in the $Z$-basis.
Using this local circuit, we define the disentangler from the SET Hamiltonian \eqref{eq: e anyon pump Hamiltonain} to the decoupled toric code as a product \eqref{eq: disentanlger C_4} over all pairs of plaquettes which are aligned horizontally.
Note that the disentangler locally preserves the onsite $\mathbb{Z}_{2}$ symmetry but does not locally commute with the $C_{4}$ symmetry since the local circuit \eqref{eq: disentanlger C_4} does not commute with the $C_4$ symmetry. 

\subsection{Parameterized family: higher-order anyon pumping}
\label{sec: anyon pump}

By interpolating the symmetry, one can introduce a parametrization:
\begin{equation}
    A_{v}^{e}(\theta)\coloneqq \adjincludegraphics[scale=1.2,trim={15pt 15pt 15pt 15pt},valign = c]{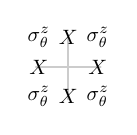},
\end{equation}
where $\sigma^{z}_{\theta}\coloneqq e^{i\frac{\theta}{4}\sigma^{x}}\sigma^{z}e^{-i\frac{\theta}{4}\sigma^{x}}$.
By using the parametrized stabilizer, we can define a parametrized family of topologically ordered systems by replacing $A_{v}^{e}$ in the Hamiltonian \eqref{eq: e anyon pump Hamiltonain} with $A_{v}^{e}(\theta)$:
\begin{equation}
    H_{\rm I}^{e}(\theta)\coloneqq H_{\rm I}^{e}\Bigm\vert_{A_{v}^{e}\mapsto A_{v}^{e}(\theta)}.
\end{equation}
Note that $\sigma^{z}_{\theta}$ itself is not $2\pi$-periodic, $H(\theta)$ is $2\pi$-periodic thanks to the $\mathbb{Z}_{2}$ onsite symmetry.

\subsection{Nontriviality of the family}\label{sec: nontriviality anyon pump}

To show that the model described above defines a nontrivial family, we construct its dual model obtained by gauging $1$-form symmetry generated by $e$-stings.
Upon gauging the $\mathbb{Z}_{2}$ $1$-form symmetry, the theory is mapped to a short-range entangled system and the dual model acquires a new (dual) $\mathbb{Z}_{2}$ $0$-form symmetry.
Therefore, by demonstrating that the model is a nontrivial SPT phase protected by the $\mathbb{Z}_{2}\times\mathbb{Z}_{2}$ and $C_{4}$ symmetries,  we can confirm the nontriviality of the original model as a family of topologically ordered systems.

\paragraph{Gauging $1$-form symmetry}
Let us briefly review the gauging procedure of $1$-form symmetry.
We first introduce a new qubit on each vertex of the lattice as an analogue of the $2$-form gauge field.
We will denote the Pauli matrices acting on these new qubits by $\tau^{x}$ and $\tau^{z}$.
Next, we impose the Gauss law constraint 
\begin{equation}
    G_{e} \coloneqq \adjincludegraphics[scale=1.25,trim={10pt 10pt 10pt 10pt},valign = c]{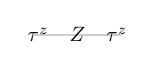}=1,\;\;
\end{equation}
on each edge and introduce minimal couplings to the Hamiltonian 
\begin{equation}
    A_{v}^{\rm gauged}\coloneq\adjincludegraphics[scale=1.25,trim={10pt 10pt 10pt 10pt},valign = c]{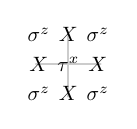},\;\;
    B_{p}^{(1)\rm gauged}\coloneq B_{p}^{(1)},\;\;
    B_{p}^{(2)\rm gauged}\coloneq B_{p}^{(2)},
\end{equation}
so that it commutes with $G_{e}$. 
The gauged Hamiltonian is given by
\begin{equation}
    H_{\mathrm{gauged}} = -\sum_{v} A_{v}^{\mathrm{gauged}} - \sum_{p} B_{p}^{(1)\mathrm{gauged}} - \sum_{p} B_{p}^{(2)\mathrm{gauged}},
\end{equation}
which is defined on the Hilbert space satisfying the Gauss law constraint $G_{e}=1$ for all edges.
By applying the following disentangler, we can decouple the original edge qubits:
\begin{equation}
    U_{\mathrm{dis}} \coloneqq \prod_{e} \mathrm{CNOT}_{e,v_{1}}\mathrm{CNOT}_{e,v_{2}},
\end{equation}
where $v_{1}$ and $v_{2}$ are the vertices connected by the edge $e$ and $e$ is the target bit.
Then, each term of the Hamiltonian is transformed as
\begin{equation}\label{eq: gauged terms}
    A_{v}^{\rm gauged} \mapsto \adjincludegraphics[scale=1.25,trim={10pt 10pt 10pt 10pt},valign = c]{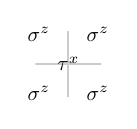},\;\;
    B_{p}^{(1)\rm gauged} \mapsto \adjincludegraphics[scale=1.25,trim={10pt 10pt 10pt 10pt},valign = c]{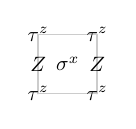},\;\;
    B_{p}^{(2)\rm gauged} \mapsto \adjincludegraphics[scale=1.25,trim={10pt 10pt 10pt 10pt},valign = c]{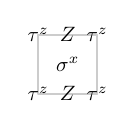},
\end{equation}
with the Gauss law constraint being transformed to $Z=1$.
Thus, we can eliminate the original edge qubits and obtain the dual Hamiltonian defined only on the vertex qubits.

\begin{figure}[t]
    \centering
    \begin{minipage}{0.32\linewidth}
        \adjincludegraphics[scale=1.5,trim={10pt 10pt 10pt 10pt},valign = c]{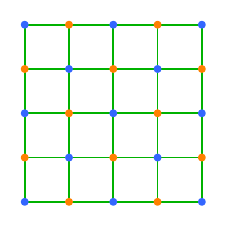}
      \caption*{(a)}
    \end{minipage}
    \hfill
    \begin{minipage}{0.32\linewidth}
        \adjincludegraphics[scale=1.5,trim={10pt 10pt 10pt 10pt},valign = c]{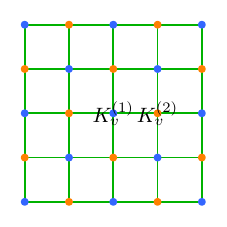}
      \caption*{(b)}
    \end{minipage}
    \hfill
    \begin{minipage}{0.32\linewidth}
        \adjincludegraphics[scale=1.5,trim={10pt 10pt 10pt 10pt},valign = c]{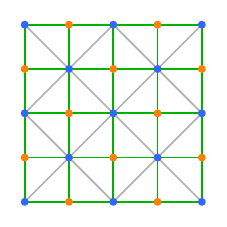}
      \caption*{(c)}
    \end{minipage}
  
    \caption{
        The lattice structure of the dual model.
        (a) The lattice where the dual model is defined.
        Qubits are placed on each vertex.
        The vertices are divided into those colored blue (where the sum of the $x$- and $y$-coordinates is even) and those colored orange (where the sum is odd).
        (b) The configuration of the stabilizers. $K_{v}^{(1)}$ is put on the type-1 (blue) vertices and $K_{v}^{(2)}$ is put on the type-2 (orange) vertices.
        (c) The relative position of the lattice to the original lattice.
        This gray lattice is the same as the one on which the toric code was defined in Section~\ref{sec: crystalline Ham}.
    }
    \label{fig: gauging_lattice}
\end{figure}

\paragraph{Crystalline SPT phase}
After eliminating the original edge qubits, the dual model \eqref{eq: gauged terms} can be recognized as the 2d cluster Hamiltonian.
For convenience of description, we rotate the model by 45 degrees and introduce a new lattice.
This lattice has a sublattice structure, as shown in Fig.~\ref{fig: gauging_lattice} (a).
We denote the vertices where the sum of the $x$- and $y$-coordinates is even by type-1 vertex (colored blue in Fig.~\ref{fig: gauging_lattice} (a)) and those where the sum is odd by type-2 vertex (colored orange in Fig.~\ref{fig: gauging_lattice} (a)).
The cluster stabilizers on each vertex are defined as follows:
\begin{equation}
    K_{v}^{(1)}\coloneq\adjincludegraphics[scale=1.25,trim={10pt 10pt 10pt 10pt},valign = c]{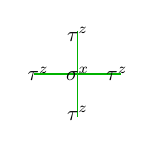},\;\;
    K_{v}^{(2)}\coloneq\adjincludegraphics[scale=1.25,trim={10pt 10pt 10pt 10pt},valign = c]{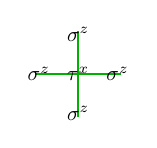},
\end{equation}
and the cluster Hamiltonian is defined as
\begin{equation}\label{eq: cluster 2d}
    H_{\mathrm{cluster}} = -\sum_{v} K_{v}^{(1)} -\sum_{v} K_{v}^{(2)}.
\end{equation}
This model has crystalline $C_{4}$-symmetry and $\mathbb{Z}_{2}\times\mathbb{Z}_2$ symmetry generated by
\begin{equation}
    U^{(1)}\coloneqq \prod_{v\in \text{type-1}} \tau^{x}_{v},\;\; U^{(2)}\coloneqq \prod_{v\in \text{type-2}} \sigma^{x}_{v},
\end{equation}
where the first (second) $\mathbb{Z}_{2}$ symmetry acts on the type-1 (type-2) vertices.
Note that $U^{(1)}$ is the dual symmetry of the $1$-form symmetry and $U^{(2)}$ is the same as the original $\mathbb{Z}_{2}$ symmetry generated by $U_{e}$.

\begin{figure}[t]
    \centering
    \begin{minipage}{0.32\linewidth}
        \adjincludegraphics[scale=1.5,trim={10pt 10pt 10pt 10pt},valign = c]{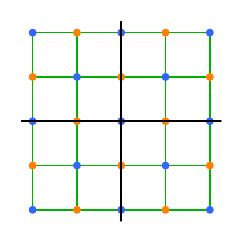}
      \caption*{(a)}
    \end{minipage}
    \hfill
    \begin{minipage}{0.32\linewidth}
        \adjincludegraphics[scale=1.5,trim={10pt 10pt 10pt 10pt},valign = c]{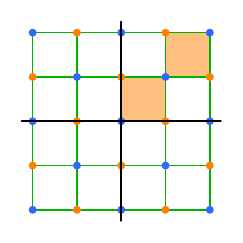}
      \caption*{(b)}
    \end{minipage}
    \hfill
    \begin{minipage}{0.32\linewidth}
        \adjincludegraphics[scale=1.5,trim={10pt 10pt 10pt 10pt},valign = c]{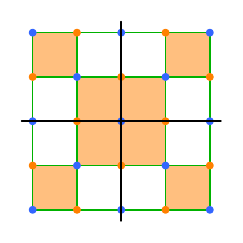}
      \caption*{(c)}
    \end{minipage}
  
    \caption{
        The procedure to show that the dual model is a nontrivial SPT phase protected by the $\mathbb{Z}_{2}\times\mathbb{Z}_{2}$ and $C_{4}$ symmetry.
        (a) Dividing the lattice into four regions by lines parallel to the $x$- and $y$-axes.
        (b) Applying the disentangler $U_{\mathrm{CZ}}$ to the first quadrant.
        The orange squares represent the four products of $\mathrm{CZ}$ operators acting on the links around the plaquette defined in d\eqref{eq: disentangler}.
        (c) Applying the disentangler to the other quadrants in a $C_{4}$ symmetric way.
        As a result, a nontrivial SPT state appears along the black lines.
    }
    \label{fig: crystalline_spt}
\end{figure}

Let us check that the dual model is a nontrivial SPT phase protected by the $\mathbb{Z}_{2}\times\mathbb{Z}_2$ and $C_{4}$ symmetry, following \cite{Song_2017}.
The procedure is as follows:
\begin{enumerate}
    \item Introduce lines parallel to the $x$- and $y$-axes to divide the lattice into four regions. (Fig.~\ref{fig: crystalline_spt} (a))
    \item Apply a disentangler that preserves the $\mathbb{Z}_{2}\times\mathbb{Z}_{2}$ symmetry on first quadrant. (Fig.~\ref{fig: crystalline_spt} (b))
    \item Apply the same disentangler to the other quadrants in a $C_{4}$ symmetric way. (Fig.~\ref{fig: crystalline_spt} (c))
    \item Verify that a nontrivial SPT state appears along the $x$- and $y$-axes.
\end{enumerate}
We use the following operator as the disentangler:
\begin{equation}\label{eq: disentangler}
    U_{\mathrm{CZ}} \coloneqq \prod_{p} \mathrm{CZ}_{v^{p}_{1}v^{p}_{2}}\mathrm{CZ}_{v^{p}_{2}v^{p}_{3}}\mathrm{CZ}_{v^{p}_{3}v^{p}_{4}}\mathrm{CZ}_{v^{p}_{4}v^{p}_{1}},
\end{equation}
where $v^{p}_{1},v^{p}_{2},v^{p}_{3},v^{p}_{4}$ are the vertices of the plaquette $p$ in clockwise order.
We denote this operator by orange squares in Fig.~\ref{fig: crystalline_spt} (b) and (c).
Each layer of $U_{\mathrm{CZ}}$ is invariant under the $\mathbb{Z}_{2}\times\mathbb{Z}_{2}$ symmetry and $U_{\mathrm{CZ}}$ disentangles the stabilizer on each plaquette in the bulk of the first quadrant as
\begin{equation}
    K^{(1)}_{v} \mapsto \sigma^{x}_{v},\;\; K^{(2)}_{v} \mapsto \tau^{x}_{v}.
\end{equation}
Thus, the qubits in the bulk of the first quadrant are disentangled to the product state.
We apply the same disentangler to the other quadrants in a $C_{4}$ symmetric way as in Fig.~\ref{fig: crystalline_spt} (c).
As a result, the stabilizers at all vertices except those lying on the $x$- and $y$-axes are mapped to either $\sigma^{x}$ or $\tau^{x}$.
On the other hand, the stabilizers at the vertices on the $x$- and $y$-axes (except for the rotation center) are transformed as
\begin{equation}
    K^{(1)}_{v} \mapsto \adjincludegraphics[scale=1.25,trim={10pt 10pt 10pt 10pt},valign = c]{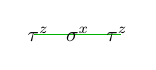},\;\; 
    K^{(2)}_{v} \mapsto \adjincludegraphics[scale=1.25,trim={10pt 10pt 10pt 10pt},valign = c]{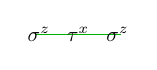},
\end{equation}
and the stabilizer at the rotation center remains unchanged.
This is precisely the one-dimensional cluster Hamiltonian and belongs to a nontrivial SPT phase protected by the $\mathbb{Z}_{2}\times\mathbb{Z}_{2}$ symmetry.
Therefore, as long as the $\mathbb{Z}_{2}\times\mathbb{Z}_{2}$ and $C_{4}$ symmetries are preserved, this model cannot be trivialized.

\paragraph{Remark on the line operator of SET state}
As mentioned in Section~\ref{sec: crystalline Ham}, the SET Hamiltonian \eqref{eq: e anyon pump Hamiltonain} has the $m$-anyon $1$-form symmetry where the $\sigma^z$ operator is attached at the corner of the line as \eqref{eq: m-anyon crystalline}. Here we discuss the origin of the attached operator on corners.
To see this, we first note that the usual $m$-anyon line operator (without corner $\sigma^z$'s) along a cycle $\gamma_{m}$ is mapped to the following surface operators under gauging the $1$-form symmetry:
\begin{equation}
    \mathcal{O}(S)\coloneqq\prod_{v\in S}\tau_{v}^x,
\end{equation}
where $S$ is a surface such that $\partial S=\gamma_{m}$. This operator is topological in the sense that its expectation value for the ground state of the trivial SPT Hamiltonian 
\begin{equation}
   H_{\mathrm{trivial}}=-\sum_{v}\sigma_{v}^x-\sum_{v}\tau_{v}^x
\end{equation}
is constant. Thus, the usual $m$-anyon operator becomes a symmetry in the gauged model of $H_{\mathrm{trivial}}$, which is the toric code, namely the trivial SET state.\footnote{For a general (non-fixed point) trivial SPT state, the decay of the expectation of $\mathcal{O}(S)$ obeys the Perimeter law, i.e., the decay is suppressed by the length of the boundary of $S$. In the ungauged toric code phase picture, this is understood by the fact that the $m$-anyon line is the charged object of the $1$-form symmetry generated by the $e$-anyon line.}
On the other hand, $\mathcal{O}(S)$ is not topological in the ground state of the cluster Hamiltonian \eqref{eq: cluster 2d}. Instead, the unitary transformed operator
\begin{equation}
    U_{\mathrm{CZ}}\mathcal{O}(S)U_{\mathrm{CZ}}^{-1}=\prod_{v\in S}\tau_{v}^x\prod_{p\in\mathrm{cor(\gamma_m)}}\sigma_{p}^{z}
\end{equation}
commutes with the cluster Hamiltonian.
Non-commutativity between $U_{\mathrm{CZ}}$ and $\mathcal{O}(S)$ reflects the fact that $U_{\mathrm{CZ}}$ is the disentangler of the crystalline SPT and does not locally commute with the symmetry operator.
The surface operator $U_{\mathrm{CZ}}\mathcal{O}(S)U_{\mathrm{CZ}}^{-1}$ is safely mapped to the line operator \eqref{eq: m-anyon crystalline} under the gauging, and so the correct $1$-form symmetry operator of the SET model \eqref{eq: e anyon pump Hamiltonain} is \eqref{eq: m-anyon crystalline}.

\subsection{Other anyon pumping}
\label{sec: other anyon pump}
Similar to the $e$-anyon pump model, we can consider $m$-anyon pumps and $f$-anyon pumps. 
To define the $m$-anyon pump, we add ancilla qubits to the vertices and consider the following stabilizers:
\begin{align}
    A_v^{m,(1)} \coloneqq  
        \vcenter{
        \hbox{
            \adjincludegraphics[scale=1.2,trim={10pt 10pt 10pt 10pt},valign = c]{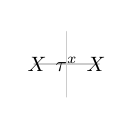}
        }
        },\;
    A_v^{m,(2)} \coloneqq  
        \vcenter{
        \hbox{
            \adjincludegraphics[scale=1.2,trim={10pt 10pt 10pt 10pt},valign = c]{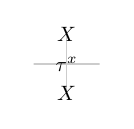}
        }
        },\;
    B_p^{m} \coloneqq  
        \vcenter{
        \hbox{
            \adjincludegraphics[scale=1.2,trim={10pt 10pt 10pt 10pt},valign = c]{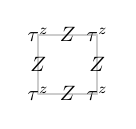}
        }
        }.
\end{align}
This model has a $\mathbb{Z}_{2}$ symmetry generated by
\begin{equation}
    U^{m}\coloneqq \prod_{v} \tau^{x}_v,
\end{equation}
and the crystalline $C_{4}$-symmetry.
Then, we can define an $m$-anyon pump model by interpolation.

\section{Summary and Future Directions}
In this paper, we have constructed various families of topologically ordered states and discussed their topological pumping properties. 
This work clarifies the physical significance of non-triviality as a family topologically ordered states realized in lattice systems.
We conclude by listing open questions and future directions.

\paragraph{Construction of Topological Invariants}

In this paper, we constructed families of topologically ordered states and investigate their nontriviality.
On the other hand, for families of topologically ordered states, the following general conjectures have been proposed:
Consider the moduli space $\mathcal{M}(\mathcal{B})$ of $2+1$ dimensional topologically ordered phases described by a modular tensor category $\mathcal{B}$. 
This space was first considered by Kitaev in Appendix F of \cite{Kitaev_2006}. 
In \cite{ENO2010}, Etingof, Nikshych and Ostrik introduced the notion of Picard $2$-group $\mathrm{Pic}(\mathcal{B})$ of $\mathcal{B}$, and investigated the homotopy type of it. It is conjectured that this algebraic structure captures part of the homotopy type of $\mathcal{M}(\mathcal{B})$.
From a physical perspective, the $3$-group $B\mathrm{Pic}(\mathcal{B})$ can be regarded as a category of invertible topological defects with various codimensions in the topologically ordered system, such that homotopy classes of paths in $\mathcal{M}(\mathcal{B})$ correspond to superselection sectors of defects obtained by varying parameters slowly in space.
Hsin, Kapustin and the last author pointed out that,  based on the cobordism hypothesis \cite{Baez:1995xq,lurie2009classificationtopologicalfieldtheories}, there is a relation between invertible defects and parametrized families in Sec.~VII. of \cite{Hsin:2020cgg},
and Hsin and Wang later delved into the relation using effective actions and lattice models \cite{Hsin:2022iug}.
These results implies the following physical interpretation of the parametrized family for 2+1d topologically ordered states:
\begin{enumerate}
    \item $S^1$-families are classified by invertible codimension-one defects.\footnote{Mathematically, they are classified by braided autoequivalences. Typically, an element is a defect that exchanges anyons, namely an anyon exchange defect, but it can also include more subtle symmetries known as soft symmetries \cite{Davydov:2013xov,Kobayashi:2025ykb}.}
    \item $S^2$-families are classified by $1$-form symmetries carried by abelian anyons.
    \item $S^3$- and $S^4$-families are expected to be captured by higher Berry phases.
\end{enumerate}
In this way, the structure of $B\mathrm{Pic}(\mathcal{B})$ provides a unifying framework for understanding the parametrized families of topologically ordered systems. 
Our analysis of nontriviality based on boundary algebras and other methods is consistent with these results.
However, a space (or homotopy type) of topologically ordered lattice models has yet to be constructed, so it remains unclear whether the models constructed in this paper realize nontriviality in the above sense. We expect that when this space is constructed, the models we propose will help to realize some of the non-trivial homotopy groups of this space. Developing a method to discuss familes of lattice models and define general topological invariants is an important open problem toward a complete classification.

\paragraph{Geometric Phases in Topologically Ordered States}

The recent establishment of geometric phases in many-body systems -- namely, the higher Berry phase -- has deepened our understanding of higher pumping phenomena in these systems.
However, in the formulations developed so far, only higher Berry phases for invertible states have essentially been formulated, and the formulation for higher Berry phases in degenerate systems, including topologically ordered states, remains incomplete.
Constructing models with higher Berry phases by deforming the toric code model is expected to provide important insight into understanding the higher Berry phases of topologically ordered states.

\paragraph{DHR bimodule}
We used the boundary algebra to demonstrate the nontriviality of the family.
For the boundary algebra $A$ (more precisely, a net of algebras), it is known that one can construct the DHR bimodule category $\mathrm{DHR}(A)$
Moreover, if $A$ satisfies weak algebraic Haag duality, there exists canonical braiding on $\mathrm{DHR}(A)$ \cite{Jones:2023ptg}.
In particular, for the toric code this is known to be equivalent, as a braided tensor category, to the Drinfeld center $Z(\mathrm{Vec}_{\mathbb{Z}_2})$ of $\mathrm{Vec}_{\mathbb{Z}_2}$.
Since the $\mathrm{DHR}$ map is known to be functorial, an $S^{1}$-family of boundary algebras corresponds to a braided autoequivalence of $\mathrm{DHR}(A)$, and an $S^{2}$-family is expected to correspond to a natural isomorphism of the identity functor on $\mathrm{DHR}(A)$.
At least, for the toric code, the Kramers-Wannier duality of the boundary algebra corresponds to the braided autoequivalence that exchanges $e$ and $m$\cite{Jones_2026,Evans:2025msy} and we have discussed its interpretation in the DHR framework above. However, we have not analyzed this interpretation for the $T^2$-family, which is a natural future direction.
Using this mathematical framework to characterize families of commuting projector Hamiltonians with more general parameter spaces, such as higher dimensional spheres, is an interesting challenge. One can consider this as a challenge of defining the homotopy type for quantum cellular automata in the boundary algebra, akin to the programs carried out in \cite{czajka2025anomalieslatticehomotopyquantum,Tu:2025bqf,ludewig2026quantumcellularautomatacoarse,ji2026quantumcellularautomatagroup}. In particular, we note that there is a good notion of blend in the boundary algebra, so the approach of \cite{czajka2025anomalieslatticehomotopyquantum} may be directly applied. We commented on the analysis of the $T^2$-family from this point of view. Likewise, we would like to know if the boundary algebra methods we employed would be useful in showing non-contractibility of families beyond those defined by commuting projector Hamiltonians.

\section*{Acknowledgements}

S.O. is supported by the European Union’s Horizon 2020 research and innovation programme through grant no. 863476 (ERC- CoG SEQUAM). 
S.O. is also supported by JSPS KAKENHI Grant Number 24K00522.
T.A.~is supported by JSPS KAKENHI Grant Number 25KJ1557.
T.A.~was also supported by JST CREST (Grant No.~JPMJCR19T2) and JSPS Overseas Challenge Program for Young Researchers. R.T. is supported by the Bhaumik Presidential Term Chair.
We thank Sasza Czajka, Sameer Erramilli, Roman Geiko, Lei Gioia, Jeongwan Haah, Jiasheng He, Corey Jones, Kyle Kawagoe, David Penneys, Shu-Heng Shao, Ruben Verresen, and Carolyn Zhang for insightful conversations and related collaborations.

\appendix

\section{Realizing symmetry actions from gauged models}
In this appendix, we explain a procedure to obtain lattice models for symmetry-enriched topologically (SET) ordered phases with general non-anomalous global symmetries.\footnote{In this appendix, we only discuss finite invertible global symmetries.}
It is believed that any RG fixed topologically ordered states, whose IR theories are TQFTs, are obtained out of (higher) braided tensor categories \cite{Kong:2014qka, Johnson-Freyd:2020usu}, and there are several ways to write lattice models for a wide class of topologically ordered phases.
For example, the string-net model and its generalizations \cite{Levin:2004mi, Kitaev:2011dxc, Lan:2013wia, Lin:2014aca, Hu:2015dga, Lake:2016mky, Hahn:2020cgf, Lin:2020bak} describe various non-chiral $2+1\mathrm{D}$ topologically ordered phases. 
By using the string-net models, one can explicitly obtain anyonic line operators that realize $1$-form symmetries of the topologically ordered states. Nevertheless, it is hard to give explicit symmetry actions of an SET phase for a given topologically ordered lattice model in general.
The reason for this comes from the fact that couplings between global symmetries and TQFTs consist of additional data, such as the symmetry fractionalization class.
Moreover, general topologically ordered states can have emergent global symmetries such as the $em$-exchange symmetry in the toric code. Such symmetry operators are obtained by condensation (higher gauging) defects in the corresponding IR TQFT \cite{Kong:2013aya, Gaiotto:2019xmp, Roumpedakis:2022aik}, and there are no systematic ways to realize such global symmetry operators explicitly.\footnote{Mathematically, the data of SET is given by a map from $BG$ of the classifying space of the symmetry group $G$ to the appropriate moduli space of the TQFT.}

Even with that difficulty, if a global symmetry of a given SET is non-anomalous, we can gauge the symmetry and may write down the lattice model for the corresponding gauged TQFTs.
According to the general theory of gauging, the gauged theory has a dual global symmetry, with which we can safely \emph{regauge}---that is, gauge the dual symmetry of a gauged theory. This operation returns the original theory: the gauged theory with respect to the dual symmetry is the same as the original theory, and the original global symmetry is equal to the other dual symmetry in the regauging.
These dual symmetry operators are often easy to write down; they act faithfully on the Hilbert space of the gauged Hamiltonian.\footnote{Although the dual symmetries act faithfully on the UV Hilbert space, they can act trivially on the IR ground states. SETs with symmetry fractionalization classes may give such examples, and more generally, symmetry transmutations are also examples of this phenomenon\cite{Seiberg:2025bqy}.} 
Since our original theory is understood as the gauged theory of dual global symmetries, we can write SET lattice models with explicit expressions for enriched global symmetry operators.

We now outline the procedure to obtain generic SET lattice models with non-anomalous global symmetries.
We first consider an SET phase, i.e., a set of a topologically ordered model and an action of a global symmetry. 
If the corresponding global symmetry is non-anomalous, one can gauge it.\footnote{To complete gauging on the lattice, we need to require the vanishing of obstruction classes, which are in general not equal to those expected from the continuous theories. See \cite{Kapustin:2025nju, Tu:2025bqf, Shirley:2025yji} for discussions of the obstruction classes in the lattice models.}
Let us assume that we have specified the TQFT for the gauged topologically ordered model and its dual symmetry action, then write the Hamiltonian of the gauged theory as $\sum_{i}h_{i}$. Then consider regauge with respect to the dual symmetry.
The regauged Hamiltonian has a form
\begin{equation}\label{eq:regauge}
    H=\sum_{i}h^{g}_{i}+\sum_{s}P^{G}_{s}+\sum_{p}P^{F}_{p},
\end{equation}
where $h_{i}^g$ is the local interaction term minimally coupled gauge fields, $P_{s}^{G}$ is a local projector onto the gauge invariant subspace, and $P_{p}^{F}$ is a local projector to ensure the flatness condition of the gauge field.
%To summarize, 
%\begin{enumerate}
    %\item Specify the gauged theory with respect to the global symmetry of the SET state. The gauged theory is in general realized by the (generalized) string-net model and the dual global symmetry 
    %\item Write down the regauged lattice Hamiltonian 
%\end{enumerate}
In the rest of this appendix, we give some examples of the procedure. 

\subsection{\texorpdfstring{$\mathbb{Z}_2$}{Z2} symmetry-enriched toric code from \texorpdfstring{$\mathcal{Z}(\mathrm{Ising})$}{Z(Ising}}
The IR TQFT of the toric code is known to be the $\mathbb{Z}_2$ gauge theory, and the $em$ exchange symmetry flows to a non-anomalous $\mathbb{Z}_2$ $0$-form symmetry, which is realized by condensing or higher gauging the fermionic line operator of the TQFT. 
If we gauge the $\mathbb{Z}_2$ symmetry, the TQFT becomes another TQFT whose anyon statistics are described by the Drinfeld center of the Ising category $\mathcal{Z}(\mathrm{Ising})$\cite{Barkeshli:2014cna}. The gauged theory has a dual $\mathbb{Z}_2$ $1$-form symmetry.
Lattice model for such a TQFT can be constructed as a string-net model. The regauged model \eqref{eq:regauge} was explicitly given in \cite{Heinrich:2016wld}.
The model is defined on a honeycomb lattice, and the $em$-exchange symmetry acts on the model as an on-site $\mathbb{Z}_2$ symmetry.

\subsection{\texorpdfstring{$\mathbb{Z}_2\times\mathbb{Z}_2$}{Z2xZ2} symmetry-enriched toric code from \texorpdfstring{$ \mathrm{D}_8$}{D8} quantum double}\label{sec:D8}
Here we discuss the toric code theory with a symmetry fractionalization that traps an anyon is obtained by regauging a $\mathbb{Z}_2\times\mathbb{Z}_2$ $1$-form symmetry in the $\mathrm{D}_8$ gauge theory.

Let us write the action of the $\mathbb{Z}_2$ gauge theory as
\begin{equation}
    \frac{1}{2}\int adb,
\end{equation}
where $a,b$ are $\mathbb{Z}_2$ cochain valued dynamical fields. Then consider a $\mathbb{Z}_2\times\mathbb{Z}_2$ $0$-form symmetry and couple its background fields $C_1,C_2$ to the action as 
\begin{equation}\label{eq:TC_Z2Z2_sym}
    \frac{1}{2}\int adb+aC_1C_2.
\end{equation}
Let us consider the gauged theory of the $\mathbb{Z}_2\times\mathbb{Z}_2$ symmetry. The action of the gauged theory is
\begin{equation}
    \frac{1}{2}\int adb+c_1db_1+c_2db_2+ac_1c_2.
\end{equation} 
Here we introduced two Lagrangian multipliers $b_1$ and $b_2$ to make $c_1$ and $c_2$ cocycles. This action realizes the action of the quantum double of $\mathrm{D}_8$, the dihedral group of order eight.
%denoted by $D(\mathrm{D}_8)$. 
%When one gauges the $\mathbb{Z}_2\times\mathbb{Z}_2$ symmetry, one obtains another TQFT called the quantum double of $\mathrm{D}_8$, the dihedral group of order eight.
To see this, we first note that $\mathrm{D}_8$ fits into the following central extension of groups:
\begin{equation}
    1 \to \mathbb{Z}_2^{a_1} \to \mathrm{D}_8 \to \mathbb{Z}_2^{a_2}\times\mathbb{Z}_2^{a_3} \to 1.
\end{equation}
Specifically, the group mulplication law of $\mathrm{D}_8$ is given by
\begin{gather}
    \begin{split}
        (a_1,(a_2,a_3))\cdot (a_1^\prime,(a_2^\prime,a_3^\prime))=(a_1+a_2+\omega((a_2,a_3),(a_2^\prime,a_3^\prime)),(a_2+a_3,a_2^\prime+a_3^\prime)),\\
    a_i,a_i^\prime\in \mathbb{Z}_2,\quad (i=1,2,3).
    \end{split}
\end{gather}
Here, $\omega$ is an element of the second cohomogoly group $\mathrm{H}^2(\mathbb{Z}_2\times\mathbb{Z}_2,\mathbb{Z}_2)\cong\mathbb{Z}_2^3$ and given as
\begin{equation}
    \omega((a_1,a_2),(a_1^\prime,a_2^\prime))=a_1a_2^\prime.
\end{equation}
Therefore, the $\mathbb{Z}_2$ gauge field for the center of $\mathrm{D}_8$ satisfies $da_1=a_2\cup a_3$. Since the partition function of the $\mathrm{D}_8$ gauge theory is given by integrating out all possible configurations of $\mathrm{D}_8$ gauge fields over the spacetime manifold, 
the partition function $Z$ can be written as 
\begin{equation}
    Z=\int\mathcal{D}a_1\mathcal{D}a_2\mathcal{D}a_3\mathcal{D}h_1\mathcal{D}h_2\mathcal{D}h_3\,
    \exp\left(\frac{2\pi i}{2}\int a_1dh_1+a_2dh_2+a_3dh_3+a_2a_3h_1\right).
\end{equation}
One can easily see that the above two actions are equivalent, i.e., the gauged theory of \eqref{eq:TC_Z2Z2_sym} is equivalent to the quantum double of $\mathrm{D}_8$. Note that the $\mathrm{D}_8$ quantum double theory has two $\mathbb{Z}_2$ $1$-form symmetries, which are generated by Wilson lines $\oint a_2,\oint a_3$, and each corresponds to a nontrivial one-dimensional irreducible representation of $\mathrm{D}_8$.

\subsubsection*{Concrete lattice model}
Here, we briefly give a concrete lattice model of the toric code phase with a nontrivial symmetry fractionalization class. The construction is based on what we saw in this subsection; regauging two $\mathbb{Z}_2$ $1$-form symmetries in the quantum double model of $\mathrm{D}_8$ \cite{Kitaev_2003}.
We define our $\mathrm{D}_8$ as 
\begin{equation}
    \mathrm{D}_8=\langle r,s\mid r^4=s^2=1, srs=r^{-1}\rangle.
\end{equation}
The local Hilbert space of the QD model is the regular representation of $\mathrm{D}_8$ and we denote the basis by $\{\ket{g}\}_{g\in\mathrm{D}_8}$.
The left and right shift operator $X_{g}$ is defined by
\begin{equation}
    \overrightarrow{X}_{g}\coloneqq  \sum_{h\in \mathrm{D}_8}\ket{gh}\bra{h},\quad 
    \overleftarrow{X}_{g}\coloneqq  \sum_{h\in \mathrm{D}_8}\ket{hg}\bra{h}.
\end{equation}
For each one-dimensional representation of $\mathrm{D}_8$, denoted by $\rho$, we define the operator $Z_{\rho}$ as
\begin{equation}
    Z_{\rho}\coloneqq  \sum_{g\in \mathrm{D}_8}\rho(g)\ket{g}\bra{g}.
\end{equation}
Speficically, we write two representations $\rho_r,\rho_s$, which are defined by 
\begin{equation}
    \rho_r(r)=-1,\rho_r(s)=1,\quad \rho_s(r)=1,\rho_s(s)=-1.
\end{equation}
To define the $\mathrm{D}_8$ quantum double model, we first define the following two operators for each vertex $v$ and plaquette $p$ as follows:\footnote{We use the notation of \cite{Cui:2019lvb}.} 
\begin{gather}
  A_v^{\mathrm{D}_8}(g) \Bigg\vert
  \begin{tikzpicture}[baseline={([yshift=-.5ex]current bounding box.center)},vertex/.style={anchor=base,circle,fill=black!25,minimum size=18pt,inner sep=2pt}]
    \draw[fill] (0,0) circle [radius=0.025] node[above right] {$v$};
    \draw (-1,0) -- (1,0);
    \draw (0,-1) -- (0,1); 
    \draw[->] (0,0) -- (-0.5,0) node[above] { $g_1$};
    \draw[->] (0,0) -- (0,-0.5) node[left] { $g_2$};    
    \draw[->] (0,0) -- (0.5,0) node[below] { $g_3$};
    \draw[->] (0,1) -- (0,0.5) node[above right] { $g_4$};
  \end{tikzpicture}
\Bigg\rangle \coloneqq
  \Bigg\vert
  \begin{tikzpicture}[baseline={([yshift=-.5ex]current bounding box.center)},vertex/.style={anchor=base,circle,fill=black!25,minimum size=18pt,inner sep=2pt}]
    \draw[fill] (0,0) circle [radius=0.025] node[above right] {$v$};
    \draw (-1,0) -- (1,0);
    \draw (0,-1) -- (0,1);
    \draw[->] (0,0) -- (-0.5,0) node[above] {$gg_1$};
    \draw[->] (0,0) -- (0,-0.5) node[left] {$gg_2$};    
    \draw[->] (0,0) -- (0.5,0) node[below] {$gg_3$};
    \draw[->] (0,1) -- (0,0.5) node[above right] {$g_4g^{-1}$};
  \end{tikzpicture}
  \Bigg\rangle,\quad 
  g,g_1,g_2,g_3,g_4\in\mathrm{D}_8,\\
  B_{p}^{\mathrm{D}_8}(h) \Bigg \vert
  \begin{tikzpicture}[baseline={([yshift=-.5ex]current bounding box.center)},vertex/.style={anchor=base,circle,fill=black!25,minimum size=18pt,inner sep=2pt}]
    \draw[fill] (0,0) circle [radius=0.025] node[below left] {$v$};
    \draw (0,0) -- (1,0) -- (1,1) -- (0,1) -- (0,0); 
    \draw[->] (0,0)--(0.5,0) node[below] {$h_1$};
    \draw[->] (1,1)--(1,0.5) node[right] {$h_2$};
    \draw[->] (1,1)--(0.5,1) node[above] {$h_3$};
    \draw[->] (0,1)--(0,0.5) node[left] {$h_4$};
    \node at (0.5,0.5) {$p$};
  \end{tikzpicture}
  \Bigg \rangle \coloneqq
  \delta_{h, h_1 h_2^{-1} h_3 h_4} 
  \Bigg \vert
  \begin{tikzpicture}[baseline={([yshift=-.5ex]current bounding box.center)},vertex/.style={anchor=base,circle,fill=black!25,minimum size=18pt,inner sep=2pt}]
    \draw[fill] (0,0) circle [radius=0.025] node[below left] {$v$};
    \draw (0,0) -- (1,0) -- (1,1) -- (0,1) -- (0,0); 
    \draw[->] (0,0)--(0.5,0) node[below] {$h_1$};
    \draw[->] (1,1)--(1,0.5) node[right] {$h_2$};
    \draw[->] (1,1)--(0.5,1) node[above] {$h_3$};
    \draw[->] (0,1)--(0,0.5) node[left] {$h_4$};
    \node at (0.5,0.5) {$p$};
  \end{tikzpicture}
  \Bigg \rangle,\quad 
  h,h_1,h_2,h_3,h_4\in\mathrm{D}_8.
  %\end{split}
\end{gather}
We further define
\begin{equation}
    A_{v}^{\mathrm{D}_8}\coloneqq  \frac{1}{|G|}\sum_{g\in \mathrm{D}_8}A_{v}^{\mathrm{D}_8}(g).
\end{equation}
Then the ground state of the $\mathrm{D}_8$ quantum double model is specified by
\begin{equation}
    A_{v}^{\mathrm{D}_8}(g)\ket{\text{GS}}=\ket{\text{GS}},\quad
    B_{p}^{\mathrm{D}_8}(h)\ket{\text{GS}}=\delta_{1,h}\ket{\text{GS}},
\end{equation}
for any $v,p,g,h$.
%Our $\mathrm{D}_8$ QD Hamiltonian is 
%\begin{equation}
    %H_{\mathrm{D}_8}=-\sum_{p}B_p^{\mathrm{D}_8}-\sum_{v}A_v^{\mathrm{D}_8}.
%\end{equation}
This $\mathrm{D}_8$ quantum double model has two $\mathbb{Z}_2$ $1$-form symmetries generated by 
\begin{equation}\label{eq:D8_1form}
    \prod_{l\in\gamma}Z_{\rho_1,l},\quad \prod_{l\in\gamma}Z_{\rho_2,l}.
\end{equation}
Before implementing the regauge, it might be useful to see how the $1$-form symmetry acts on local shift operators.
One can see that 
\begin{equation}
    Z_{\rho_r}\overrightarrow{X}_{r}=-\overrightarrow{X}_{r}Z_{\rho_r},\quad Z_{\rho_r}\overrightarrow{X}_{s}=\overrightarrow{X}_{s}Z_{\rho_r}.
\end{equation}
Indeed,
\begin{equation}
    Z_{\rho_r}\overrightarrow{X}_{r}\ket{r^{i}s^{j}}=Z_{\rho_r}\ket{r^{i+1}s^{j}}=(-1)^{i+1}\ket{r^{i+1}s^{j}}=-\overrightarrow{X}_{r}Z_{\rho_r}\ket{r^{i}s^{j}},
\end{equation}
and the latter equality can be shown from the fact that 
\begin{equation*}
    \overrightarrow{X}_{s}\ket{r^{i}s^{j}}=
    \begin{cases*}
        \ket{r^{i}s^{j+1}} & if $i$ is even, \\
        \ket{r^{i+2}s^{j+1}} & if $i$ is odd.
    \end{cases*}
    =\ket{r^{i+1+(-1)^{i+1}}s^{j+1}}
\end{equation*}
Similarly, one can find 
\begin{equation}
    Z_{\rho_s}\overrightarrow{X}_{s}=-\overrightarrow{X}_{s}Z_{\rho_s},\quad Z_{\rho_s}\overrightarrow{X}_{r}=\overrightarrow{X}_{r}Z_{\rho_s}.
\end{equation}
Upon $\mathbb{Z}_2\times\mathbb{Z}_2$ gauging, we have following symmetry operators:
\begin{equation}
    U_{r}\coloneqq  \prod_{v}\sigma^x_{v,r},\quad U_{s}\coloneqq  \prod_{v}\sigma^x_{v,s}.%\quad
    %U_{rs}\coloneqq  \prod_{v}\sigma^x_{v,r}\sigma^x_{v,s},
\end{equation}
Now let us choose generators of our $\mathbb{Z}_2\times\mathbb{Z}_2$ $0$-form symmetry as
\begin{equation}\label{eq:TC_sym_regauge}
    U_1=U_{s},\quad U_{2}=U_{r}U_{s}=\prod_{v}\sigma^x_{v,r}\sigma^x_{v,s}.
\end{equation}
Consider the action of these symmetry operators on the right-half infinite plane. Since the boundary truncation is not unique, we take the truncation $U_{1}^{R},U_{2}^{R}$ as 
\begin{equation}
    U_{1}^{R}\ket{\text{GS}}=\prod_{l}\overrightarrow{X}_{s,l}\ket{\text{GS}},\quad U_{2}^{R}\ket{\text{GS}}=\prod_{l}\overrightarrow{X}_{rs,l}\ket{\text{GS}},\quad
    (U_{1}U_{2})^{R}\ket{\text{GS}}=\prod_{l}\overrightarrow{X}_{r,l}\ket{\text{GS}},
\end{equation}
where $l$ runs over all links along the boundary.
Then we see that 
\begin{equation}
    U_{1}^{R}U_{2}^{R}\ket{\text{GS}}=\left(\prod_{l}\overrightarrow{X}_{r^2,l}\right)U_{2}^{R}U_{1}^{R}\ket{\text{GS}}=\left(\prod_{l}\overrightarrow{X}_{r^2,l}\right)(U_{1}U_{2})^{R}\ket{\text{GS}}.
\end{equation}
Since $\prod_{l}\overrightarrow{X}_{r^2,l}$ commutes two $1$-form symmetries \eqref{eq:D8_1form}, it is still a symmetry operator in the regauged theory and corresponds to an abelian anyon in the toric code. Therefore, the $\mathbb{Z}_2\times\mathbb{Z}_2$ $0$-form symmetry is coupled to the IR action of the toric code as \eqref{eq:TC_Z2Z2_sym} up to a relabeling $a$ and $b$. Using this regauged model with the symmetry \eqref{eq:TC_sym_regauge}, one can construct a nontrivial $2$-parameter family by interpolating the symmetry.

\bibliography{bibliography}

\providecommand{\href}[2]{#2}\begingroup\raggedright\begin{thebibliography}{100}

\bibitem{Thouless83}
D.J.~Thouless, \emph{Quantization of particle transport}, \href{https://doi.org/10.1103/PhysRevB.27.6083}{\emph{Phys. Rev. B} {\bfseries 27} (1983) 6083}.

\bibitem{niu1984quantised}
Q.~Niu and D.~Thouless, \emph{Quantised adiabatic charge transport in the presence of substrate disorder and many-body interaction}, \href{https://doi.org/10.1088/0305-4470/17/12/016}{\emph{{J. Phys. A: Math. Gen.}} {\bfseries 17} (1984) 2453}.

\bibitem{Teo:2010zb}
J.C.Y.~Teo and C.L.~Kane, \emph{{Topological Defects and Gapless Modes in Insulators and Superconductors}}, \href{https://doi.org/10.1103/PhysRevB.82.115120}{\emph{Phys. Rev. B} {\bfseries 82} (2010) 115120} [\href{https://arxiv.org/abs/1006.0690}{{\ttfamily 1006.0690}}].

\bibitem{Cordova:2019jnf}
C.~C\'ordova, D.S.~Freed, H.T.~Lam and N.~Seiberg, \emph{{Anomalies in the Space of Coupling Constants and Their Dynamical Applications I}}, \href{https://doi.org/10.21468/SciPostPhys.8.1.001}{\emph{SciPost Phys.} {\bfseries 8} (2020) 001} [\href{https://arxiv.org/abs/1905.09315}{{\ttfamily 1905.09315}}].

\bibitem{Cordova:2019uob}
C.~C\'ordova, D.S.~Freed, H.T.~Lam and N.~Seiberg, \emph{{Anomalies in the Space of Coupling Constants and Their Dynamical Applications II}}, \href{https://doi.org/10.21468/SciPostPhys.8.1.002}{\emph{SciPost Phys.} {\bfseries 8} (2020) 002} [\href{https://arxiv.org/abs/1905.13361}{{\ttfamily 1905.13361}}].

\bibitem{Hsin:2020cgg}
P.-S.~Hsin, A.~Kapustin and R.~Thorngren, \emph{{Berry Phase in Quantum Field Theory: Diabolical Points and Boundary Phenomena}}, \href{https://doi.org/10.1103/PhysRevB.102.245113}{\emph{Phys. Rev. B} {\bfseries 102} (2020) 245113} [\href{https://arxiv.org/abs/2004.10758}{{\ttfamily 2004.10758}}].

\bibitem{Tantivasadakarn:2021wdv}
N.~Tantivasadakarn, R.~Thorngren, A.~Vishwanath and R.~Verresen, \emph{{Pivot Hamiltonians as generators of symmetry and entanglement}}, \href{https://doi.org/10.21468/SciPostPhys.14.2.012}{\emph{SciPost Phys.} {\bfseries 14} (2023) 012} [\href{https://arxiv.org/abs/2110.07599}{{\ttfamily 2110.07599}}].

\bibitem{Jones:2024zhx}
N.G.~Jones, A.~Prakash and P.~Fendley, \emph{{Pivoting through the chiral-clock family}}, \href{https://doi.org/10.21468/SciPostPhys.18.3.094}{\emph{SciPost Phys.} {\bfseries 18} (2025) 094} [\href{https://arxiv.org/abs/2406.01680}{{\ttfamily 2406.01680}}].

\bibitem{Prakash:2024yfr}
A.~Prakash and S.A.~Parameswaran, \emph{{Charge pumps, boundary modes, and the necessity of unnecessary criticality}}, \href{https://doi.org/10.1103/kztj-jcyc}{\emph{Phys. Rev. B} {\bfseries 112} (2025) L241117} [\href{https://arxiv.org/abs/2408.15351}{{\ttfamily 2408.15351}}].

\bibitem{Jones:2025khc}
N.G.~Jones, R.~Thorngren, R.~Verresen and A.~Prakash, \emph{{Charge pumps, pivot Hamiltonians, and symmetry-protected topological phases}}, \href{https://doi.org/10.1103/rtq1-pplf}{\emph{Phys. Rev. B} {\bfseries 112} (2025) 165123} [\href{https://arxiv.org/abs/2507.00995}{{\ttfamily 2507.00995}}].

\bibitem{Manjunath:2026ezp}
N.~Manjunath and D.V.~Else, \emph{{In search of diabolical critical points}},  \href{https://arxiv.org/abs/2601.10783}{{\ttfamily 2601.10783}}.

\bibitem{kitaev2015homotopy}
A.~Kitaev, ``{Homotopy-theoretic approach to SPT phases in action: \( \mathbb{Z}_{16} \) classification of three-dimensional superconductors}.'' \href{https://www.ipam.ucla.edu/abstract/?tid=12389}{Talk at the Institute for Pure and Applied Mathematics (IPAM), UCLA}, 2015.

\bibitem{Else:2016hyb}
D.V.~Else and C.~Nayak, \emph{{Classification of topological phases in periodically driven interacting systems}}, \href{https://doi.org/10.1103/PhysRevB.93.201103}{\emph{Phys. Rev. B} {\bfseries 93} (2016) 201103} [\href{https://arxiv.org/abs/1602.04804}{{\ttfamily 1602.04804}}].

\bibitem{roy2017floquet}
R.~Roy and F.~Harper, \emph{Floquet topological phases with symmetry in all dimensions}, \href{https://doi.org/10.1103/PhysRevB.95.195128}{\emph{Phys. Rev. B} {\bfseries 95} (2017) 195128} [\href{https://arxiv.org/abs/1610.06899}{{\ttfamily 1610.06899}}].

\bibitem{Tantivasadakarn:2021noi}
N.~Tantivasadakarn and A.~Vishwanath, \emph{{Symmetric Finite-Time Preparation of Cluster States via Quantum Pumps}}, \href{https://doi.org/10.1103/PhysRevLett.129.090501}{\emph{Phys. Rev. Lett.} {\bfseries 129} (2022) 090501} [\href{https://arxiv.org/abs/2107.04019}{{\ttfamily 2107.04019}}].

\bibitem{Inamura:2026hjl}
K.~Inamura and S.~Ohyama, \emph{{Generalized cluster states in 2+1d: non-invertible symmetries, interfaces, and parameterized families}},  \href{https://arxiv.org/abs/2601.08615}{{\ttfamily 2601.08615}}.

\bibitem{Kapustin:2020mkl}
A.~Kapustin and L.~Spodyneiko, \emph{{Higher-dimensional generalizations of the {T}houless charge pump}},  \href{https://arxiv.org/abs/2003.09519}{{\ttfamily 2003.09519}}.

\bibitem{Shiozaki:2021weu}
K.~Shiozaki, \emph{{Adiabatic cycles of quantum spin systems}}, \href{https://doi.org/10.1103/PhysRevB.106.125108}{\emph{Phys. Rev. B} {\bfseries 106} (2022) 125108} [\href{https://arxiv.org/abs/2110.10665}{{\ttfamily 2110.10665}}].

\bibitem{Artymowicz:2023erv}
A.~Artymowicz, A.~Kapustin and N.~Sopenko, \emph{{Quantization of the Higher Berry Curvature and the Higher Thouless Pump}}, \href{https://doi.org/10.1007/s00220-024-05026-2}{\emph{Commun. Math. Phys.} {\bfseries 405} (2024) 191} [\href{https://arxiv.org/abs/2305.06399}{{\ttfamily 2305.06399}}].

\bibitem{Ohyama:2026oay}
S.~Ohyama and K.~Inamura, \emph{{Parameterized families of 2+1d $G$-cluster states}},  \href{https://arxiv.org/abs/2601.08616}{{\ttfamily 2601.08616}}.

\bibitem{Kitaev_Freed60}
A.Y.~Kitaev, \emph{Differential forms on the space of statistical mechanical lattice models}, {\emph{talk at Between Topology and Quantum Field Theory: a conference in celebration of Dan Freed’s 60th birthday} (2019) }.

\bibitem{Kapustin:2020eby}
A.~Kapustin and L.~Spodyneiko, \emph{{Higher-dimensional generalizations of {B}erry curvature}}, \href{https://doi.org/10.1103/PhysRevB.101.235130}{\emph{Phys. Rev. B} {\bfseries 101} (2020) 235130} [\href{https://arxiv.org/abs/2001.03454}{{\ttfamily 2001.03454}}].

\bibitem{Wen:2021gwc}
X.~Wen, M.~Qi, A.~Beaudry, J.~Moreno, M.J.~Pflaum, D.~Spiegel et~al., \emph{{Flow of higher Berry curvature and bulk-boundary correspondence in parametrized quantum systems}}, \href{https://doi.org/10.1103/PhysRevB.108.125147}{\emph{Phys. Rev. B} {\bfseries 108} (2023) 125147} [\href{https://arxiv.org/abs/2112.07748}{{\ttfamily 2112.07748}}].

\bibitem{Ohyama:2023suc}
S.~Ohyama, Y.~Terashima and K.~Shiozaki, \emph{{Discrete higher Berry phases and matrix product states}}, \href{https://doi.org/10.1103/PhysRevB.110.035114}{\emph{Phys. Rev. B} {\bfseries 110} (2024) 035114} [\href{https://arxiv.org/abs/2303.04252}{{\ttfamily 2303.04252}}].

\bibitem{OR23}
S.~Ohyama and S.~Ryu, \emph{{Higher structures in matrix product states}}, \href{https://doi.org/10.1103/PhysRevB.109.115152}{\emph{Phys. Rev. B} {\bfseries 109} (2024) 115152} [\href{https://arxiv.org/abs/2304.05356}{{\ttfamily 2304.05356}}].

\bibitem{Qi:2023ysw}
M.~Qi, D.T.~Stephen, X.~Wen, D.~Spiegel, M.J.~Pflaum, A.~Beaudry et~al., \emph{{Charting the space of ground states with tensor networks}}, \href{https://doi.org/10.21468/SciPostPhys.18.5.168}{\emph{SciPost Phys.} {\bfseries 18} (2025) 168} [\href{https://arxiv.org/abs/2305.07700}{{\ttfamily 2305.07700}}].

\bibitem{Spiegel:2023lhv}
D.D.~Spiegel, \emph{{A C*-Algebraic Approach to Parametrized Quantum Spin Systems and Their Phases in One Spatial Dimension}}, Ph.D. thesis, Colorado U., PHYSICS, 2023.
\newblock \href{https://arxiv.org/abs/2305.07951}{{\ttfamily 2305.07951}}.

\bibitem{shiozaki2023higher}
K.~Shiozaki, N.~Heinsdorf and S.~Ohyama, \emph{{Higher Berry curvature from matrix product states}}, \href{https://doi.org/10.1103/1cbf-kxny}{\emph{Phys. Rev. B} {\bfseries 112} (2025) 035154} [\href{https://arxiv.org/abs/2305.08109}{{\ttfamily 2305.08109}}].

\bibitem{Ohyama:2024jsg}
S.~Ohyama and S.~Ryu, \emph{{Higher Berry connection for matrix product states}}, \href{https://doi.org/10.1103/PhysRevB.111.035121}{\emph{Phys. Rev. B} {\bfseries 111} (2025) 035121} [\href{https://arxiv.org/abs/2405.05327}{{\ttfamily 2405.05327}}].

\bibitem{Ohyama:2024ytt}
S.~Ohyama and S.~Ryu, \emph{{Higher Berry phase from projected entangled pair states in (2+1) dimensions}}, \href{https://doi.org/10.1103/PhysRevB.111.045112}{\emph{Phys. Rev. B} {\bfseries 111} (2025) 045112} [\href{https://arxiv.org/abs/2405.05325}{{\ttfamily 2405.05325}}].

\bibitem{Sommer:2024dtb}
O.E.~Sommer, X.~Wen and A.~Vishwanath, \emph{{Higher Berry Curvature from the Wave Function. I. Schmidt Decomposition and Matrix Product States}}, \href{https://doi.org/10.1103/PhysRevLett.134.146601}{\emph{Phys. Rev. Lett.} {\bfseries 134} (2025) 146601} [\href{https://arxiv.org/abs/2405.05316}{{\ttfamily 2405.05316}}].

\bibitem{Sommer:2024lzp}
O.E.~Sommer, A.~Vishwanath and X.~Wen, \emph{{Higher Berry curvature from the wave function. II. Locally parametrized states beyond one dimension}}, \href{https://doi.org/10.1103/PhysRevB.111.155110}{\emph{Phys. Rev. B} {\bfseries 111} (2025) 155110} [\href{https://arxiv.org/abs/2405.05323}{{\ttfamily 2405.05323}}].

\bibitem{Liu:2024ulq}
B.~Liu, J.~Zhang, S.~Ohyama, Y.~Kusuki and S.~Ryu, \emph{{Multiwavefunction overlap and multientropy for topological ground states in (2+1) dimensions}}, \href{https://doi.org/10.1103/tjcf-yryh}{\emph{Phys. Rev. B} {\bfseries 112} (2025) 125160} [\href{https://arxiv.org/abs/2410.08284}{{\ttfamily 2410.08284}}].

\bibitem{Geiko:2024cra}
R.~Geiko, \emph{{Parametrized topological phases in 1d and T-duality}},  \href{https://arxiv.org/abs/2412.20905}{{\ttfamily 2412.20905}}.

\bibitem{Spiegel:2025ugu}
D.D.~Spiegel, M.~Qi, D.T.~Stephen, M.~Hermele, M.J.~Pflaum and A.~Beaudry, \emph{{A Classifying Space for Phases of Matrix Product States}}, \href{https://doi.org/10.1007/s00220-025-05491-3}{\emph{Commun. Math. Phys.} {\bfseries 407} (2026) 17} [\href{https://arxiv.org/abs/2501.14241}{{\ttfamily 2501.14241}}].

\bibitem{wen2025spaceconformalboundaryconditions}
X.~Wen, \emph{Space of conformal boundary conditions from the view of higher berry phase: Flow of berry curvature in parametrized bcfts},  \href{https://arxiv.org/abs/2507.12546}{{\ttfamily 2507.12546}}.

\bibitem{Shiozaki:2025pyo}
K.~Shiozaki, \emph{{Equivariant parameter families of spin chains: A discrete MPS formulation}}, \href{https://doi.org/10.21468/SciPostPhys.20.1.024}{\emph{SciPost Phys.} {\bfseries 20} (2026) 024} [\href{https://arxiv.org/abs/2507.19932}{{\ttfamily 2507.19932}}].

\bibitem{Choi:2022odr}
Y.~Choi and K.~Ohmori, \emph{{Higher Berry phase of fermions and index theorem}}, \href{https://doi.org/10.1007/JHEP09(2022)022}{\emph{JHEP} {\bfseries 09} (2022) 022} [\href{https://arxiv.org/abs/2205.02188}{{\ttfamily 2205.02188}}].

\bibitem{Choi:2025ebk}
Y.~Choi, H.~Ha, D.~Kim, Y.~Kusuki, S.~Ohyama and S.~Ryu, \emph{{Higher Structures on Boundary Conformal Manifolds: Higher Berry Phase and Boundary Conformal Field Theory}},  \href{https://arxiv.org/abs/2507.12525}{{\ttfamily 2507.12525}}.

\bibitem{Kapustin:2022apy}
A.~Kapustin and N.~Sopenko, \emph{{Local Noether theorem for quantum lattice systems and topological invariants of gapped states}}, \href{https://doi.org/10.1063/5.0085964}{\emph{J. Math. Phys.} {\bfseries 63} (2022) 091903} [\href{https://arxiv.org/abs/2201.01327}{{\ttfamily 2201.01327}}].

\bibitem{PhysRevLett.48.1559}
D.C.~Tsui, H.L.~Stormer and A.C.~Gossard, \emph{Two-dimensional magnetotransport in the extreme quantum limit}, \href{https://doi.org/10.1103/PhysRevLett.48.1559}{\emph{Phys. Rev. Lett.} {\bfseries 48} (1982) 1559}.

\bibitem{Wen:1989iv}
X.G.~Wen, \emph{{Topological Order in Rigid States}}, \href{https://doi.org/10.1142/S0217979290000139}{\emph{Int. J. Mod. Phys. B} {\bfseries 4} (1990) 239}.

\bibitem{MooreRead1991}
G.W.~Moore and N.~Read, \emph{{Nonabelions in the fractional quantum Hall effect}}, \href{https://doi.org/10.1016/0550-3213(91)90407-O}{\emph{Nucl. Phys. B} {\bfseries 360} (1991) 362}.

\bibitem{Freedman:2000gwh}
M.H.~Freedman, M.~Larsen and Z.~Wang, \emph{{A Modular Functor Which is Universal for Quantum Computation}}, \href{https://doi.org/10.1007/s002200200645}{\emph{Commun. Math. Phys.} {\bfseries 227} (2002) 605} [\href{https://arxiv.org/abs/quant-ph/0001108}{{\ttfamily quant-ph/0001108}}].

\bibitem{Kitaev_2003}
A.~Kitaev, \emph{Fault-tolerant quantum computation by anyons}, \href{https://doi.org/10.1016/s0003-4916(02)00018-0}{\emph{Annals of Physics} {\bfseries 303} (2003) 2–30} [\href{https://arxiv.org/abs/quant-ph/9707021}{{\ttfamily quant-ph/9707021}}].

\bibitem{freedman2002topologicalquantumcomputation}
M.H.~Freedman, A.~Kitaev, M.J.~Larsen and Z.~Wang, \emph{Topological quantum computation},  \href{https://arxiv.org/abs/quant-ph/0101025}{{\ttfamily quant-ph/0101025}}.

\bibitem{BakalovKirillov}
B.~Bakalov and A.~Kirillov, \emph{Lectures on tensor categories and modular functors}, {\emph{Amer. Math. Soc. Univ. Lect. Ser.} {\bfseries 21} (2001) }.

\bibitem{Turaev2016}
V.G.~Turaev, \emph{Quantum Invariants of Knots and 3-Manifolds}, vol.~18 of \emph{de Gruyter Studies in Mathematics}, Walter de Gruyter, Berlin, 2nd~ed. (2016).

\bibitem{Kitaev_2006}
A.~Kitaev, \emph{Anyons in an exactly solved model and beyond}, \href{https://doi.org/10.1016/j.aop.2005.10.005}{\emph{Annals of Physics} {\bfseries 321} (2006) } [\href{https://arxiv.org/abs/cond-mat/0506438}{{\ttfamily cond-mat/0506438}}].

\bibitem{Aasen:2022cdu}
D.~Aasen, Z.~Wang and M.B.~Hastings, \emph{{Adiabatic paths of Hamiltonians, symmetries of topological order, and automorphism codes}}, \href{https://doi.org/10.1103/PhysRevB.106.085122}{\emph{Phys. Rev. B} {\bfseries 106} (2022) 085122} [\href{https://arxiv.org/abs/2203.11137}{{\ttfamily 2203.11137}}].

\bibitem{Hsin:2022iug}
P.-S.~Hsin and Z.~Wang, \emph{{On topology of the moduli space of gapped Hamiltonians for topological phases}}, \href{https://doi.org/10.1063/5.0136906}{\emph{J. Math. Phys.} {\bfseries 64} (2023) 041901} [\href{https://arxiv.org/abs/2211.16535}{{\ttfamily 2211.16535}}].

\bibitem{ENO2010}
{P. Etingof, D. Nikshych, V. Ostrik, with an appendix by E. Meir}, \emph{{Fusion categories and homotopy theory}}, {\emph{Quantum topology} {\bfseries 1} (2010) 209} [\href{https://arxiv.org/abs/0909.3140}{{\ttfamily 0909.3140}}].

\bibitem{Dennis_2002}
E.~Dennis, A.~Kitaev, A.~Landahl and J.~Preskill, \emph{Topological quantum memory}, \href{https://doi.org/10.1063/1.1499754}{\emph{J. Math. Phys.} {\bfseries 43} (2002) 4452–4505} [\href{https://arxiv.org/abs/quant-ph/0110143}{{\ttfamily quant-ph/0110143}}].

\bibitem{Po_2017}
H.C.~Po, L.~Fidkowski, A.~Vishwanath and A.C.~Potter, \emph{Radical chiral floquet phases in a periodically driven kitaev model and beyond}, \href{https://doi.org/10.1103/physrevb.96.245116}{\emph{Phys. Rev. B} {\bfseries 96} (2017) } [\href{https://arxiv.org/abs/1701.01440}{{\ttfamily 1701.01440}}].

\bibitem{Heinrich:2016wld}
C.~Heinrich, F.~Burnell, L.~Fidkowski and M.~Levin, \emph{{Symmetry enriched string-nets: Exactly solvable models for SET phases}}, \href{https://doi.org/10.1103/PhysRevB.94.235136}{\emph{Phys. Rev. B} {\bfseries 94} (2016) 235136} [\href{https://arxiv.org/abs/1606.07816}{{\ttfamily 1606.07816}}].

\bibitem{ma2024quantumcellularautomatasymmetric}
R.~Ma, Y.~Li and M.~Cheng, \emph{{Quantum Cellular Automata on Symmetric Subalgebras}},  \href{https://arxiv.org/abs/2411.19280}{{\ttfamily 2411.19280}}.

\bibitem{Jones_2026}
C.~Jones, K.~Schatz and D.J.~Williamson, \emph{{Quantum Cellular Automata and Categorical Dualities of Spin Chains}}, \href{https://doi.org/10.1007/s00220-026-05571-y}{\emph{Commun. Math. Phys.} {\bfseries 407} (2026) 66} [\href{https://arxiv.org/abs/2410.08884}{{\ttfamily 2410.08884}}].

\bibitem{Barkeshli:2022wuz}
M.~Barkeshli, Y.-A.~Chen, S.-J.~Huang, R.~Kobayashi, N.~Tantivasadakarn and G.~Zhu, \emph{{Codimension-2 defects and higher symmetries in (3+1)D topological phases}}, \href{https://doi.org/10.21468/SciPostPhys.14.4.065}{\emph{SciPost Phys.} {\bfseries 14} (2023) 065} [\href{https://arxiv.org/abs/2208.07367}{{\ttfamily 2208.07367}}].

\bibitem{Tu:2025bqf}
Y.-T.~Tu, D.M.~Long and D.V.~Else, \emph{{Anomalies of Global Symmetries on the Lattice}}, \href{https://doi.org/10.1103/m188-w1ct}{\emph{Phys. Rev. X} {\bfseries 16} (2026) 011027} [\href{https://arxiv.org/abs/2507.21209}{{\ttfamily 2507.21209}}].

\bibitem{Shirley:2025yji}
W.~Shirley, C.~Zhang, W.~Ji and M.~Levin, \emph{{Anomaly-free symmetries with obstructions to gauging and onsiteability}},  \href{https://arxiv.org/abs/2507.21267}{{\ttfamily 2507.21267}}.

\bibitem{jones2023localtopologicalorderboundary}
C.~Jones, P.~Naaijkens, D.~Penneys and D.~Wallick, \emph{{Local topological order and boundary algebras}},  \href{https://arxiv.org/abs/2307.12552}{{\ttfamily 2307.12552}}.

\bibitem{Jones:2023imy}
C.~Jones and J.~Lim, \emph{{An Index for Quantum Cellular Automata on Fusion Spin Chains}}, \href{https://doi.org/10.1007/s00023-024-01429-y}{\emph{Annales Henri Poincare} {\bfseries 25} (2024) 4399} [\href{https://arxiv.org/abs/2309.10961}{{\ttfamily 2309.10961}}].

\bibitem{Jones:2023ptg}
C.~Jones, \emph{{DHR bimodules of quasi-local algebras and symmetric quantum cellular automata}},  \href{https://arxiv.org/abs/2304.00068}{{\ttfamily 2304.00068}}.

\bibitem{Hataishi2025}
L.~Hataishi, D.~Jaklitsch, C.~Jones and M.~Yamashita, \emph{On the structure of {DHR} bimodules of abstract spin chains},  \href{https://arxiv.org/abs/2504.06094}{{\ttfamily 2504.06094}}.

\bibitem{Jones:2025yhx}
C.~Jones, P.~Naaijkens and D.~Penneys, \emph{{Holography for bulk-boundary local topological order}},  \href{https://arxiv.org/abs/2506.19969}{{\ttfamily 2506.19969}}.

\bibitem{Jones:2026dcb}
C.~Jones and X.~Yang, \emph{{On the structure of categorical duality operators}},  \href{https://arxiv.org/abs/2603.09949}{{\ttfamily 2603.09949}}.

\bibitem{Barkeshli:2014cna}
M.~Barkeshli, P.~Bonderson, M.~Cheng and Z.~Wang, \emph{{Symmetry Fractionalization, Defects, and Gauging of Topological Phases}}, \href{https://doi.org/10.1103/PhysRevB.100.115147}{\emph{Phys. Rev. B} {\bfseries 100} (2019) 115147} [\href{https://arxiv.org/abs/1410.4540}{{\ttfamily 1410.4540}}].

\bibitem{Kitaev:2011dxc}
A.~Kitaev and L.~Kong, \emph{{Models for Gapped Boundaries and Domain Walls}}, \href{https://doi.org/10.1007/s00220-012-1500-5}{\emph{Commun. Math. Phys.} {\bfseries 313} (2012) 351} [\href{https://arxiv.org/abs/1104.5047}{{\ttfamily 1104.5047}}].

\bibitem{Bombin_2010}
H.~Bombin, \emph{Topological order with a twist: Ising anyons from an abelian model}, \href{https://doi.org/10.1103/physrevlett.105.030403}{\emph{Phys. Rev. Lett.} {\bfseries 105} (2010) } [\href{https://arxiv.org/abs/1004.1838}{{\ttfamily 1004.1838}}].

\bibitem{Bridgeman:2017gyy}
J.C.~Bridgeman, S.D.~Bartlett and A.C.~Doherty, \emph{{Tensor networks with a twist: Anyon-permuting domain walls and defects in projected entangled pair states}}, \href{https://doi.org/10.1103/PhysRevB.96.245122}{\emph{Phys. Rev. B} {\bfseries 96} (2017) 245122} [\href{https://arxiv.org/abs/1708.08930}{{\ttfamily 1708.08930}}].

\bibitem{Tantivasadakarn:2023zov}
N.~Tantivasadakarn and X.~Chen, \emph{{String operators for Cheshire strings in topological phases}}, \href{https://doi.org/10.1103/PhysRevB.109.165149}{\emph{Phys. Rev. B} {\bfseries 109} (2024) 165149} [\href{https://arxiv.org/abs/2307.03180}{{\ttfamily 2307.03180}}].

\bibitem{Vanhove:2024lmj}
R.~Vanhove, V.~Ravindran, D.T.~Stephen, X.-G.~Wen and X.~Chen, \emph{{Duality via sequential quantum circuit in the topological holography formalism}}, \href{https://doi.org/10.1103/y9cd-dz5k}{\emph{Phys. Rev. B} {\bfseries 112} (2025) 035173} [\href{https://arxiv.org/abs/2409.06647}{{\ttfamily 2409.06647}}].

\bibitem{li2024domainwallssptsewing}
Y.~Li, Z.~Song, A.~Kubica and I.H.~Kim, \emph{Domain walls from spt-sewing},  \href{https://arxiv.org/abs/2411.11967}{{\ttfamily 2411.11967}}.

\bibitem{Bravyi:1998sy}
S.B.~Bravyi and A.Y.~Kitaev, \emph{{Quantum codes on a lattice with boundary}},  \href{https://arxiv.org/abs/quant-ph/9811052}{{\ttfamily quant-ph/9811052}}.

\bibitem{Chatterjee:2022kxb}
A.~Chatterjee and X.-G.~Wen, \emph{{Symmetry as a shadow of topological order and a derivation of topological holographic principle}}, \href{https://doi.org/10.1103/PhysRevB.107.155136}{\emph{Phys. Rev. B} {\bfseries 107} (2023) 155136} [\href{https://arxiv.org/abs/2203.03596}{{\ttfamily 2203.03596}}].

\bibitem{Inamura:2023ldn}
K.~Inamura and X.-G.~Wen, \emph{{2+1D symmetry-topological-order from local symmetric operators in 1+1D}},  \href{https://arxiv.org/abs/2310.05790}{{\ttfamily 2310.05790}}.

\bibitem{czajka2025anomalieslatticehomotopyquantum}
A.M.~Czajka, R.~Geiko and R.~Thorngren, \emph{{Anomalies on the Lattice, Homotopy of Quantum Cellular Automata, and a Spectrum of Invertible States}},  \href{https://arxiv.org/abs/2512.02105}{{\ttfamily 2512.02105}}.

\bibitem{Song_2017}
H.~Song, S.-J.~Huang, L.~Fu and M.~Hermele, \emph{Topological phases protected by point group symmetry}, \href{https://doi.org/10.1103/physrevx.7.011020}{\emph{Physical Review X} {\bfseries 7} (2017) } [\href{https://arxiv.org/abs/1604.08151}{{\ttfamily 1604.08151}}].

\bibitem{Baez:1995xq}
J.C.~Baez and J.~Dolan, \emph{{Higher dimensional algebra and topological quantum field theory}}, \href{https://doi.org/10.1063/1.531236}{\emph{J. Math. Phys.} {\bfseries 36} (1995) 6073} [\href{https://arxiv.org/abs/q-alg/9503002}{{\ttfamily q-alg/9503002}}].

\bibitem{lurie2009classificationtopologicalfieldtheories}
J.~Lurie, \emph{On the classification of topological field theories},  \href{https://arxiv.org/abs/0905.0465}{{\ttfamily 0905.0465}}.

\bibitem{Davydov:2013xov}
A.~Davydov, \emph{{Bogomolov multiplier, double class-preserving automorphisms, and modular invariants for orbifolds}}, \href{https://doi.org/10.1063/1.4895764}{\emph{J. Math. Phys.} {\bfseries 55} (2014) 092305} [\href{https://arxiv.org/abs/1312.7466}{{\ttfamily 1312.7466}}].

\bibitem{Kobayashi:2025ykb}
R.~Kobayashi and M.~Barkeshli, \emph{{Soft symmetries of topological orders}}, \href{https://doi.org/10.1103/cwn4-jl57}{\emph{Phys. Rev. B} {\bfseries 113} (2026) 115150} [\href{https://arxiv.org/abs/2501.03314}{{\ttfamily 2501.03314}}].

\bibitem{Evans:2025msy}
D.E.~Evans and C.~Jones, \emph{{An operator algebraic approach to fusion category symmetry on the lattice}},  \href{https://arxiv.org/abs/2507.05185}{{\ttfamily 2507.05185}}.

\bibitem{ludewig2026quantumcellularautomatacoarse}
M.~Ludewig, \emph{Quantum cellular automata are a coarse homology theory},  \href{https://arxiv.org/abs/2603.10501}{{\ttfamily 2603.10501}}.

\bibitem{ji2026quantumcellularautomatagroup}
M.~Ji and B.~Yang, \emph{Quantum cellular automata: The group, the space, and the spectrum},  \href{https://arxiv.org/abs/2602.16572}{{\ttfamily 2602.16572}}.

\bibitem{Kong:2014qka}
L.~Kong and X.-G.~Wen, \emph{{Braided fusion categories, gravitational anomalies, and the mathematical framework for topological orders in any dimensions}},  \href{https://arxiv.org/abs/1405.5858}{{\ttfamily 1405.5858}}.

\bibitem{Johnson-Freyd:2020usu}
T.~Johnson-Freyd, \emph{{On the Classification of Topological Orders}}, \href{https://doi.org/10.1007/s00220-022-04380-3}{\emph{Commun. Math. Phys.} {\bfseries 393} (2022) 989} [\href{https://arxiv.org/abs/2003.06663}{{\ttfamily 2003.06663}}].

\bibitem{Levin:2004mi}
M.A.~Levin and X.-G.~Wen, \emph{{String net condensation: A Physical mechanism for topological phases}}, \href{https://doi.org/10.1103/PhysRevB.71.045110}{\emph{Phys. Rev. B} {\bfseries 71} (2005) 045110} [\href{https://arxiv.org/abs/cond-mat/0404617}{{\ttfamily cond-mat/0404617}}].

\bibitem{Lan:2013wia}
T.~Lan and X.-G.~Wen, \emph{{Topological quasiparticles and the holographic bulk-edge relation in (2+1) -dimensional string-net models}}, \href{https://doi.org/10.1103/PhysRevB.90.115119}{\emph{Phys. Rev. B} {\bfseries 90} (2014) 115119} [\href{https://arxiv.org/abs/1311.1784}{{\ttfamily 1311.1784}}].

\bibitem{Lin:2014aca}
C.-H.~Lin and M.~Levin, \emph{{Generalizations and limitations of string-net models}}, \href{https://doi.org/10.1103/PhysRevB.89.195130}{\emph{Phys. Rev. B} {\bfseries 89} (2014) 195130} [\href{https://arxiv.org/abs/1402.4081}{{\ttfamily 1402.4081}}].

\bibitem{Hu:2015dga}
Y.~Hu, N.~Geer and Y.-S.~Wu, \emph{{Full dyon excitation spectrum in extended Levin-Wen models}}, \href{https://doi.org/10.1103/PhysRevB.97.195154}{\emph{Phys. Rev. B} {\bfseries 97} (2018) 195154} [\href{https://arxiv.org/abs/1502.03433}{{\ttfamily 1502.03433}}].

\bibitem{Lake:2016mky}
E.~Lake and Y.-S.~Wu, \emph{{Signatures of broken parity and time reversal symmetry in generalized string-net models}}, \href{https://doi.org/10.1103/PhysRevB.94.115139}{\emph{Phys. Rev. B} {\bfseries 94} (2016) 115139} [\href{https://arxiv.org/abs/1605.07194}{{\ttfamily 1605.07194}}].

\bibitem{Hahn:2020cgf}
A.~Hahn and R.~Wolf, \emph{{Generalized string-net model for unitary fusion categories without tetrahedral symmetry}}, \href{https://doi.org/10.1103/PhysRevB.102.115154}{\emph{Phys. Rev. B} {\bfseries 102} (2020) 115154} [\href{https://arxiv.org/abs/2004.07045}{{\ttfamily 2004.07045}}].

\bibitem{Lin:2020bak}
C.-H.~Lin, M.~Levin and F.J.~Burnell, \emph{{Generalized string-net models: A thorough exposition}}, \href{https://doi.org/10.1103/PhysRevB.103.195155}{\emph{Phys. Rev. B} {\bfseries 103} (2021) 195155} [\href{https://arxiv.org/abs/2012.14424}{{\ttfamily 2012.14424}}].

\bibitem{Kong:2013aya}
L.~Kong, \emph{{Anyon condensation and tensor categories}}, \href{https://doi.org/10.1016/j.nuclphysb.2014.07.003}{\emph{Nucl. Phys. B} {\bfseries 886} (2014) 436} [\href{https://arxiv.org/abs/1307.8244}{{\ttfamily 1307.8244}}].

\bibitem{Gaiotto:2019xmp}
D.~Gaiotto and T.~Johnson-Freyd, \emph{{Condensations in higher categories}},  \href{https://arxiv.org/abs/1905.09566}{{\ttfamily 1905.09566}}.

\bibitem{Roumpedakis:2022aik}
K.~Roumpedakis, S.~Seifnashri and S.-H.~Shao, \emph{{Higher Gauging and Non-invertible Condensation Defects}}, \href{https://doi.org/10.1007/s00220-023-04706-9}{\emph{Commun. Math. Phys.} {\bfseries 401} (2023) 3043} [\href{https://arxiv.org/abs/2204.02407}{{\ttfamily 2204.02407}}].

\bibitem{Seiberg:2025bqy}
N.~Seiberg and S.~Seifnashri, \emph{{Symmetry transmutation and anomaly matching}}, \href{https://doi.org/10.1007/JHEP09(2025)014}{\emph{JHEP} {\bfseries 09} (2025) 014} [\href{https://arxiv.org/abs/2505.08618}{{\ttfamily 2505.08618}}].

\bibitem{Kapustin:2025nju}
A.~Kapustin and S.~Xu, \emph{{Higher symmetries and anomalies in quantum lattice systems}},  \href{https://arxiv.org/abs/2505.04719}{{\ttfamily 2505.04719}}.

\bibitem{Cui:2019lvb}
S.X.~Cui, D.~Ding, X.~Han, G.~Penington, D.~Ranard, B.C.~Rayhaun et~al., \emph{{Kitaev's quantum double model as an error correcting code}}, \href{https://doi.org/10.22331/q-2020-09-24-331}{\emph{Quantum} {\bfseries 4} (2020) 331} [\href{https://arxiv.org/abs/1908.02829}{{\ttfamily 1908.02829}}].

\end{thebibliography}\endgroup

\end{document}